\documentclass[twocolumn,amsmath,preprintnumbers,superscriptaddress,amssymb]{revtex4}
\usepackage{amssymb}
\usepackage{makeidx}
\usepackage{amsmath}
\usepackage{appendix}
\usepackage{dcolumn}
\usepackage{bm}
\usepackage{graphicx}
\usepackage{soul}
\usepackage{amsfonts}
\usepackage{color}
\usepackage{psfrag}

\begin{document}
\title{Control of spectroscopic features of multiphoton
transitions in two coupled qubits by driving fields}

\author{V. O. Munyaev}
\author{M. V. Bastrakova}
\email{bastrakova@phys.unn.ru}

\affiliation{Nizhny Novgorod State University, 23 Gagarin Ave., 603022, Nizhny Novgorod, Russia}

\begin{abstract}
The quantum levels population behavior of the two coupled flux qubits depending on the external driving field characteristics is studied. The explicit expressions for the multiphoton transition probabilities at an arbitrary control field amplitude is obtained for the case of small tunnel splitting energies. We describe the controllable features of their formation and thereby creating or destroying entanglement by system bias tuning on the direct inter-level transition and during the transition through intermediate states. We found a feature of the qubits population inverting that ends in the independence of the resonances positions from the qubits coupling strength. Using Floquet--Markov equation we numerically demonstrate, that the positions of multiphoton resonances are stable to dissipative processes. 

\end{abstract}

\maketitle

\section{Introduction}

Initially, the study of high-order harmonic generation phenomena, weak localization, parametric amplification, Raman scattering, frequency conversion and mixing was carried out on ensembles of natural atoms \cite{Vahala2004}. These developments have found important scientific and practical applications in the field of quantum optoelectronic technology. Recently there has been an increase in interest in the study of such effects in artificial atoms, in which the control of energy levels (energy spectra) is more flexible and is carried out by adjusting external parameters. One of the most promising and rapidly developing artificial systems is superconducting quantum circuits with Josephson junctions \cite{Wendin2017, Kockum2019, Kjaergaard2020}.

By present time, the technique of packing superconducting circuits into 2D or 3D structures, coupling circuits with  transmission lines and incorporating them into the resonators with high quality factor is created an developed \cite{Oelsner2010,Zagoskin2011,Gu2017}. Recently the first algorithmic quantum computer \cite{Wang2018,Reagor2018,Neill2018} based on superconducting qubits was created and a quantum supremacy in solving a limited range of mathematical problems was demonstrated \cite{Arute2019}. Superconducting devices are strongly connected to the electromagnetic field, that opens up an opportunity to observe a wide range of interesting nonlinear effects in the microwave range \cite{Gu2017}, such as multiphoton transitions and Landau--Zener--St\"{u}ckelberg interference \cite{Landau1932,Zener1932,Stuckelberg1932} characterized by the emergence of Floquet states \cite{Shirley1965}. Microwave driving fields with variable amplitude and fixed frequency has become a tool to analyze quantum coherence and to access the energy structure of coupled qubits under the strong driving \cite{Shevchenko2010}. At the low-frequency limit, the system evolves mainly adiabatically during the field period, except for small time intervals when the levels experience quasicrossing and quantum coherent Landau--Zener tunneling can be observed. This multiphoton  spectroscopy technique has been  experimentally investigated in superconducting flux qubits \cite{Oliver2005,Berns2006,Rudner2008,Izmalkov2008,Neilinger2016}, charge qubit \cite{Sillanpaa2006}, and quantum dots devices \cite{Ribeiro2013,Kohler2018}. In addition, the Landau--Zener interference can be used to determine relevant information about the qubit coupling with a noisy environment \cite{Blattmann2015, Kohler2018}, as well as to implement protocols for controllable entanglement in quantum tomography of qubit states \cite{Quintana2013,Roch2014}.

In this paper, we study multiphoton transitions in a system of two coupled flux qubits placed in a strong driving magnetic field. We have theoretically studied in detail the processes of the multiphoton resonances occurrence in the framework of the Floquet perturbation theory, based on  the small  qubit  tunneling  splitting. The obtained analytical results are compared with the numerical analysis in the framework of the quasi-energy representation. This concept allows us to explain the formation of the quantum coherent tunneling regions when controlling the bias of qubits \cite{Ilichev2010}. As a result of the developed theory, the principles of the formation of the inverse population of the excited state with respect to the ground state were analytically explained. In addition, we investigate the stability of the detected multiphoton effects relative to the dissipation effects and study the entanglement formation in the system.

The paper is organized as follows. In Sec. II, we describe the periodically driven system of two coupled qubits using the Floquet formalism. In Sec. III, we propose a method for calculating the probabilities of multiphoton transitions between the states of interacting qubits within the framework of the Floquet perturbation theory. In Sec. IV, we study the transition probabilities near the found conditions of multiphoton resonances within the framework of the rotating wave approximation of the Floquet state. The multiphoton interference patterns formation based on the developed theory and numerical calculation in the framework of the Floquet--Markov equation are presented in Sec. V. In Sec. VI we discuss the discovered spectroscopic features of multiphoton effects and make a conclusions.

\section{SYSTEM MODEL UNDER STUDY AND THE FLOQUET FORMALISM}

The investigated system consists of two coupled superconducting flux qubits \cite{Majer2005,Weber2017}. The system are described by the global Hamiltonian:

\begin{equation}
\begin{gathered}
\hat{H}\left(t\right)=-\frac{1}{2}\sum_{q=1}^{2}\left(\epsilon_q\sigma_z^{\left(q\right)}+\Delta_q\sigma_x^{\left(q\right)}\right)+\hat{H}_{g}+\hat{V}\left(t\right),
\end{gathered}
\label{eq:hamiltonian_global}
\end{equation}
where $\sigma_z^{\left(q\right)}$, $\sigma_x^{\left(q\right)}$ are the Pauli matrices, with $q=1,2$ the index of each qubit. The energy bias, $\epsilon_q$, can be controlled experimentally with a magnetic flux in the case of flux qubits, the tunnel level splitting, $\Delta_q$, is  fixed device parameters and determined by the relation among the charging energy and the Josephson energy of the junctions in the circuit.  The term, $\hat{H}_{g}$,  in Eq.~(1)  is the coupling Hamiltonian, which in the case of flux qubits can be written as:
\begin{equation}
\begin{gathered}
\hat{H}_{g}=-\frac{g}{2}\sigma_z^{\left(1\right)}\sigma_z^{\left(2\right)},
\end{gathered}
\label{eq:hamiltonian_coupling}
\end{equation}
where the parameter $g$  quantifies the strength of the interaction between the qubits. The ferromagnetic / antiferromagnetic interaction
between qubits can be implemented using an additional superconducting circuit \cite{Weber2017}, which effectively affects the coupling strength $g=\pm~|g|$.  In the presence of the external driving fields we have the term
\begin{equation}
\begin{gathered}
\hat{V}\left(t\right)=-\frac{1}{2}\sum_{q=1}^{2}v_{q}\left(t\right)\sigma_z^{\left(q\right)},
\end{gathered}
\label{eq:hamiltonian_driving}
\end{equation}
where $v_{q}\left(t\right)=A_{q}\cos\left(\omega t-\varphi_0\right)$ is a time-periodic magnetic flux of the microwave amplitude field, $A_{q}$, and frequency, $\omega$, applied to each qubit. We assume that the system is under the influence of a pulse sequence of an alternating fields with fixed period $T=2\pi/\omega$. At the same time, there may occur losses and phase shifts of the pulses when passing through the coaxial lines, which will affect the arrival time of the pulse on the qubits. To account for this effect, let’s denote the random time, $t_0$, of the pulse arrival on the qubits or the corresponding random phase $\varphi_0=t_0\omega$.

The resulting Hamiltonian is periodic in time, $\hat{H}\left(t\right)=\hat{H}\left(t+T\right)$. According to the Floquet theorem \cite{Shirley1965, Grifoni1998}, the solution of the Schrödinger equation can be spanned in the Floquet basis $\left\{\left|u_\alpha\left(t\right)\right>\right\}$ as $\left|\varPsi\left(t\right)\right>=\sum_{\alpha}c_{\alpha} e^{-i\gamma_{\alpha} t}\left|u_{\alpha}\left(t\right)\right>$, with $\gamma_{\alpha}$ are the quasi-energies \cite{Zeldovich1967,Ritus1967,Sambe1973} ($\alpha$ is the quantum number determining the
quasi-energy and for two coupled qubits $\alpha = 1,2,3,4$). The coefficients $c_{\alpha}$ are defined from the initial condition $\left|\varPsi\left(t_0\right)\right>$ at time $t_0$. The time-evolution for Floquet states is given by
\begin{equation}
\begin{gathered}
\left(\hat{H}\left(t\right)-i\frac{\partial}{\partial t}\right)\left|u_{\alpha}\left(t\right)\right>=\gamma_{\alpha}\left|u_{\alpha}\left(t\right)\right>,
\end{gathered}
\label{eq:equation_floquet}
\end{equation}
and they satisfy $\left|u_{\alpha}\left(t+T\right)\right>=\left|u_{\alpha}\left(t\right)\right>$. By solving Eq.~\eqref{eq:equation_floquet}, we can find the time-independent eigenvalues belonging to the Floquet zone $\gamma_{\alpha}\in\left[ -\omega/2, \omega/2 \right]$. 

It is useful to characterize the state transition of a time dependent system with the average probability:
\begin{equation}
\begin{gathered}
\bar{P}_{a\to b}=\lim_{\tau\to\infty}\frac{1}{\tau}\int_{0}^{\tau}\text{d} t\frac{1}{T} \int_{0}^{T}\text{d} t_0 P_{a\to b}\left(t,t_0\right),
\end{gathered}
\label{eq:average_probability_definition}
\end{equation}
where $P_{a\to b}\left(t,t_0\right)$ is the probability of transition from state $\left|a\right>$ to state $\left|b\right>$ at time $t$ with initial phase $\varphi_0$. The transition probability described by the expression~\eqref{eq:average_probability_definition} is a function of the duration of the microwave external field $\tau=t-t_0$.

Consider the system Eq.~\eqref{eq:hamiltonian_global} with a zero initial phase ($\varphi_0=0$) and denote its Floquet eigenstates and quasi-energies, respectively, as $\left|u_\alpha\left(t\right)\right>$ and $\gamma_{\alpha}$. Then for an arbitrary initial phase $\varphi_0$ the transition probability $P_{a\to b}\left(t,t_0\right)$, expressed through the introduced basis $\left\{\left|u_\alpha\left(t\right)\right>\right\}$, has the form
\begin{equation}
\begin{gathered}
P_{a\to b}\left(t,t_0\right)\!=\!\left|\sum_{\alpha}\!\left<b|u_{\alpha}\!\left(t\!-\!t_0\right)\right>\!e^{-i\gamma_{\alpha}t}\!\left<u_{\alpha}\!\left(-t_0\right)|a\right>\right|^2\!.
\end{gathered}
\label{eq:probability_t_phi0}
\end{equation}
Using Fourier series expansion for the Floquet eigenstates $\left|u_{\alpha}\left(t\right)\right>=\sum_{k}e^{ik\omega t}\left|u_{\alpha k}\right>$ and averaging with Eq.~\eqref{eq:average_probability_definition} for the case $\left(\gamma_{\alpha}-\gamma_{\beta\ne\alpha}\right)/\omega\notin \mathbb{Z}$ we obtain
\begin{equation}
\begin{gathered}
\bar{P}_{a\to b}=\left(S^T S\right)_{ab},
\end{gathered}
\label{eq:average_probability}
\end{equation}
where $S_{\alpha x}=\sum_{k}\left|\left<u_{\alpha k}|x\right>\right|^2$ are the elements of the matrix $S$ or equivalently, according to Parseval's identity \cite{Hazewinkel1991}, $S_{\alpha x}=\frac{1}{T}\int_{0}^{T}\text{d} t\left|\left<u_{\alpha}\left(t\right)|x\right>\right|^2$. It is useful to note that the matrix elements $S_{\alpha x}$ satisfy the equalities $\sum_{\alpha}S_{\alpha x}=1$ and $\sum_{x}S_{\alpha x}=1$ ($\left\{\left|x\right>\right\}$ is complete orthonormal basis), so matrix $S$ has only 9 independent elements.

\section{FLOQUET Perturbation Theory}\label{sec:floquet}

In this section, we will focus on the analytical study of the multiphoton nonlinear processes in the superconducting flux qubits under the influence of the strong external fields, when driving amplitudes are equal, $A_1=A_2=A$, and the controlled energy parameters exceed the tunneling splitting of qubit levels, $\Delta_{q} \ll A, \epsilon_{q}$. This condition for parameters is valid for the modern experiments on the study of Landau--Zener tunneling and amplitude spectroscopy of qubit states \cite{Berns2006,Rudner2008,Izmalkov2008,Shevchenko2010}.

According to the assumptions made, this allows us to use the perturbation theory to find the Floquet states $\left|u_{\alpha}\left(t\right)\right>$ and quasi-energies  $\gamma_{\alpha}$. Then the global Hamiltonian Eq.~\eqref{eq:hamiltonian_global} can be divided into an unperturbed part $\hat{H}_0\left(t\right)$ and a perturbing term $\hat{H}_1$
\begin{equation}
\begin{gathered}
\hat{H}\left(t\right)=\hat{H}_0\left(t\right)+\hat{H}_1,
\end{gathered}
\label{eq:hamiltonian_perturbation_division}
\end{equation}
where
\begin{gather}
\hat{H}_0\left(t\right)=-\frac{1}{2}\sum_{q=1}^{2}\epsilon_q\sigma_z^{\left(q\right)}+\hat{H}_{g}+\hat{V}\left(t\right),\label{eq:hamiltonian_unperturbed_part}\\
\hat{H}_1=-\frac{1}{2}\sum_{q=1}^{2}\Delta_q\sigma_x^{\left(q\right)}.
\label{eq:hamiltonian_perturbed_part}
\end{gather}
A Floquet equation \eqref{eq:equation_floquet} solution is sought in the form of perturbation-series in powers of smallnes parameters $\Delta_q$
\begin{equation}
\begin{aligned}
\left|u_{\alpha}\left(t\right)\right>&=\left|u_{\alpha}^{\left(0\right)}\left(t\right)\right>+\sum_{n=1}^{+\infty}\left|u_{\alpha}^{\left(n\right)}\left(t\right)\right>,\\
\gamma_{\alpha}&=\gamma_{\alpha}^{\left(0\right)}+\sum_{n=1}^{+\infty}\gamma_{\alpha}^{\left(n\right)}.
\end{aligned}
\label{eq:floquet_series_solution}
\end{equation}
The perturbative corrections $\left|u_{\alpha}^{\left(n\right)}\left(t\right)\right>$ and $\gamma_{\alpha}^{\left(n\right)}$ are defined iteratively by equations
\begin{align}
\left(\hat{H}_0\left(t\right)-i\frac{\partial}{\partial t}\right)\left|u_{\alpha}^{\left(0\right)}\left(t\right)\right>=\gamma_{\alpha}^{\left(0\right)}&\left|u_{\alpha}^{\left(0\right)}\left(t\right)\right>,\nonumber\\
\left(\hat{H}_0\left(t\right)-i\frac{\partial}{\partial t}\right)\left|u_{\alpha}^{\left(n\right)}\left(t\right)\right>=\gamma_{\alpha}^{\left(0\right)}&\left|u_{\alpha}^{\left(n\right)}\left(t\right)\right> - \label{eq:floquet_equation_unperturbed}\\
\hat{H}_1\left|u_{\alpha}^{\left(n-1\right)}\left(t\right)\right>+\sum_{m=1}^{n}\gamma_{\alpha}^{\left(m\right)}&\left|u_{\alpha}^{\left(n-m\right)}\left(t\right)\right>,\quad n\ge 1.\nonumber
\end{align}
The time-evolution operator $\hat{U}_{\alpha}\left(t\right)$ for the unperturbed system Eq.~\eqref{eq:floquet_equation_unperturbed} is
\begin{align}
\hat{U}_{\alpha}&\left(t\right)\!=\!\exp\left\{i\:\mathrm{diag}\left[\left(\frac{\epsilon_1+\epsilon_2+g}{2}\!+\!\gamma_{\alpha}^{\left(0\right)}\right)t\!+\!\frac{A}{\omega}\sin\omega t,\right.\right.\nonumber\\
&\left.\left.\left(\frac{\epsilon_1-\epsilon_2-g}{2}+\gamma_{\alpha}^{\left(0\right)}\right)t, -\left(\frac{\epsilon_1-\epsilon_2+g}{2}-\gamma_{\alpha}^{\left(0\right)}\right)t,\right.\right.\nonumber\\
&\left.\left.-\left(\frac{\epsilon_1+\epsilon_2-g}{2}\!-\!\gamma_{\alpha}^{\left(0\right)}\right)t\!-\!\frac{A}{\omega}\sin\omega t\right]\right\}.\label{eq:floquet_evolution_operator_unperturbed}
\end{align}
In order for the Floquet state $\left|u_{\alpha}\left(t\right)\right>$ to be periodic, it is necessary to require the periodicity of some evolution operator elements $\hat{U}_{\alpha}\left(t\right)$ (see Eq.~\eqref{eq:floquet_evolution_operator_unperturbed}). Further, we will consider the case of parameters $\epsilon_1$, $\epsilon_2$ and $g$ satisfying the conditions $\left(\epsilon_{1,2}\pm g\right)/\omega\notin\mathbb{Z}$ and $\left(\epsilon_{1}\pm\epsilon_{2}\right)/\omega\notin\mathbb{Z}$ (non-resonant case); consequently, all quasienergies $\gamma_{\alpha}^{\left(0\right)}$ are different and only one component of Eq.~\eqref{eq:floquet_evolution_operator_unperturbed} is $T$-periodic. Floquet states and quasi-energies$\pmod{\omega}$ in zeroth order are, respectively,
\begin{equation}
\hspace*{-0.26cm}
\begin{aligned}
&\left|u_{1}^{\left(0\right)}\!\left(t\right)\right>\!=\!\Big(\!e^{i\frac{A}{\omega}\sin\omega t},0,0,0\!\Big)^T\!,&&\gamma_{1}^{\left(0\right)}\!=\!-\frac{\epsilon_{1}\!+\!\epsilon_{2}\!+\!g}{2},\\
&\left|u_{2}^{\left(0\right)}\!\left(t\right)\right>\!=\!\Big(\!0,1,0,0\!\Big)^T\!,&&\gamma_{2}^{\left(0\right)}\!=\!-\frac{\epsilon_{1}\!-\!\epsilon_{2}\!-\!g}{2},\\
&\left|u_{3}^{\left(0\right)}\!\left(t\right)\right>\!=\!\Big(\!0,0,1,0\!\Big)^T\!,&&\gamma_{3}^{\left(0\right)}\!=\!\frac{\epsilon_{1}\!-\!\epsilon_{2}\!+\!g}{2},\\
&\left|u_{4}^{\left(0\right)}\!\left(t\right)\right>\!=\!\Big(\!0,0,0,e^{-i\frac{A}{\omega}\sin\omega t}\!\Big)^T\!,&&\gamma_{4}^{\left(0\right)}\!=\!\frac{\epsilon_{1}\!+\!\epsilon_{2}\!-\!g}{2}.
\end{aligned} \label{eq:floquet_solution_unperturbed}
\end{equation}
The general solution to the equation Eq.~\eqref{eq:floquet_equation_unperturbed} is given by
\begin{align}
&\left|u_{\alpha}^{\left(n\right)}\left(t\right)\right>=\hat{U}_{\alpha}\left(t\right)\Bigg[\left|u_{\alpha}^{\left(n\right)}\left(0\right)\right>+i\int_{0}^{t}\text{d} t'\hat{U}_{\alpha}^{\dag}\left(t'\right)\nonumber\times\\
&\!\left(\sum_{m=1}^{n}\gamma_{\alpha}^{\left(m\right)}\left|u_{\alpha}^{\left(n-m\right)}\left(t'\right)\right>-\hat{H}_1\left|u_{\alpha}^{\left(n-1\right)}\left(t'\right)\right>\right)\Bigg].
\label{eq:floquet_solution_perturbed}
\end{align}
The integration constant $\left|u_{\alpha}^{\left(n\right)}\left(0\right)\right>$ and the correction to the quasienergy $\gamma_{\alpha}^{\left(n\right)}$ are determined from the $T$-periodicity condition for the correction $\left|u_{\alpha}^{\left(n\right)}\left(t\right)\right>$, thus $\gamma_{\alpha}^{\left(n\right)}$ is chosen so integrand does not contain a constant terms. In our special non-resonant case of diagonal operator Eq.~\eqref{eq:floquet_evolution_operator_unperturbed}, we have
\begin{align}
&\qquad\left|u_{\alpha}\left(t\right)\right>=C_{\alpha}\left|u_{\alpha}^{\left(0\right)}\left(t\right)\right>+\sum_{n=1}^{+\infty}\left|u_{\alpha}^{\left(n\right)}\left(t\right)\right>,\nonumber\\
&\qquad\left|u_{\alpha}^{\left(n\right)}\left(t\right)\right>=i\hat{U}_{\alpha}\left(t\right)\int^{t}\text{d} t'\hat{U}_{\alpha}^{\dag}\left(t'\right) \times\label{eq:floquet_solution_special}\\
&\!\left(\sum_{m=1}^{n}\gamma_{\alpha}^{\left(m\right)}\left|u_{\alpha}^{\left(n-m\right)}\left(t'\right)\right>-\hat{H}_1\left|u_{\alpha}^{\left(n-1\right)}\left(t'\right)\right>\right),\quad n\ge1,\nonumber
\end{align}
where the integration constants are assumed to be zero; unperturbed solutions $\left|u_{\alpha}^{\left(0\right)}\left(t\right)\right>$ and $\gamma_{\alpha}^{\left(0\right)}$ are defined by Eq.~\eqref{eq:floquet_solution_unperturbed}. Because of the phase arbitrariness the constant $C_{\alpha}$ can be chosen real and found from the normalization, $\left\|u_{\alpha}\left(t\right)\right\|=1$. Using Eqs.~\eqref{eq:floquet_solution_unperturbed} and \eqref{eq:floquet_solution_special}, up to second order of tunnel splitting energies $\Delta_{q}$ we obtain Fourier components, $\left|u_{\alpha k}\right>$, of the Floquet states (see the explicit form of the expression in the Appendix) and quasi-energies $\gamma_{\alpha}$
\begin{equation}
\begin{aligned}
\gamma_{1}\!&=\!-\frac{\epsilon_{1}+\epsilon_{2}+g}{2}-\frac{1}{2}\left(\Delta_1^2\chi_{10}^{+}+\Delta_2^2\chi_{20}^{+}\right),\\
\gamma_{2}\!&=\!-\frac{\epsilon_{1}-\epsilon_{2}-g}{2}-\frac{1}{2}\left(\Delta_1^2\chi_{10}^{-}-\Delta_2^2\chi_{20}^{+}\right),\\
\gamma_{3}\!&=\!\frac{\epsilon_{1}-\epsilon_{2}+g}{2}+\frac{1}{2}\left(\Delta_1^2\chi_{10}^{+}-\Delta_2^2\chi_{20}^{-}\right),\\
\gamma_{4}\!&=\!\frac{\epsilon_{1}+\epsilon_{2}-g}{2}+\frac{1}{2}\left(\Delta_1^2\chi_{10}^{-}+\Delta_2^2\chi_{20}^{-}\right).
\end{aligned}
\label{eq:floquet_gamma}
\end{equation}
Here the constant $\chi_{ik}^{\pm}$ is given by expressions
\begin{equation}
\begin{gathered}
\lambda_{qk}^{\pm}=\frac{J_{\pm k}\left(A/\omega\right)}{2\left(\pm\epsilon_q+g+k\omega\right)},\\
\chi_{qk}^{\pm}=\pm\sum_{n=-\infty}^{+\infty}J_{\pm\left(n+k\right)}\left(A/\omega\right)\lambda_{qn}^{\pm},
\end{gathered}
\label{eq:lambda_chi}
\end{equation}
and for $m\ne 0$ satisfy the equalities
$\sum_{n}\lambda_{qn}^{\pm}\lambda_{q,n-m}^{\pm}=\sum_{n}\chi_{qn}^{\pm}\chi_{q,n-m}^{\pm}=\pm\frac{1}{2m\omega}\left(\chi_{qm}^{\pm}-\chi_{q,-m}^{\pm}\right)$. Here $J_z$ denote the Bessel function of the first kind.

Using the found expressions for the $\left|u_{\alpha k}\right>$ and $\gamma_{\alpha}$, we determine the transitions amplitudes between stationary levels (elements of the matrix $S$ in Eq.~(7), see the Appendix). Finally, from Eq.~\eqref{eq:average_probability} average transition probabilities $\bar{P}_{1\to 2}$, $\bar{P}_{1\to 3}$ and $\bar{P}_{1\to 4}$ are

\begin{widetext}
\begin{equation}
\begin{aligned}
\bar{P}_{1\to 2}&=\frac{\Delta_2^2}{2\epsilon_2^2}\sum_{k=-\infty}^{+\infty}\left[J_{k}\left(\frac{A}{\omega}\right)\frac{g\!+\!k\omega}{\epsilon_2\!+\!g\!+\!k\omega}\right]^2+\ldots,\\
\bar{P}_{1\to 3}&=\frac{\Delta_1^2}{2\epsilon_1^2}\sum_{k=-\infty}^{+\infty}\left[J_{k}\left(\frac{A}{\omega}\right)\frac{g\!+\!k\omega}{\epsilon_1\!+\!g\!+\!k\omega}\right]^2+\ldots,\\
\bar{P}_{1\to 4}&=
\frac{\Delta_1^2\Delta_2^2}{16\epsilon_1^2\epsilon_2^2}\sum_{k=-\infty}^{+\infty}\left\{\left[J_{k}\left(\frac{A}{\omega}\right)\frac{g\!+\!k\omega}{\epsilon_2\!+\!g\!+\!k\omega}\right]^2
\sum_{n=-\infty}^{+\infty}\left[J_{n}\left(\frac{A}{\omega}\right)\frac{g\!+\!n\omega}{-\epsilon_1\!+\!g\!+\!n\omega}\right]^2\right.\\
&\left.+\left[J_{k}\left(\frac{A}{\omega}\right)\frac{g\!+\!k\omega}{\epsilon_1\!+\!g\!+\!k\omega}\right]^2
\sum_{n=-\infty}^{+\infty}\left[J_{n}\left(\frac{A}{\omega}\right)\frac{g\!+\!n\omega}{-\epsilon_2\!+\!g\!+\!n\omega}\right]^2\right.\\
&\left.+\left[\sum_{n=-\infty}^{+\infty}J_{n}\left(\frac{A}{\omega}\right)J_{k-n}\left(\frac{A}{\omega}\right)\left(1\!-\!\frac{1}{\epsilon_1\!+\!\epsilon_2\!+\!k\omega}\left(\frac{\epsilon_1\!\left(\epsilon_1\!+\!k\omega\right)}{\epsilon_1\!+\!g\!+\!n\omega}\!+\!\frac{\epsilon_2\!\left(\epsilon_2\!+\!k\omega\right)}{\epsilon_2\!+\!g\!+\!n\omega}\right)\right)\right]^2\right.\\
&\left.+\left[\sum_{n=-\infty}^{+\infty}J_{n}\left(\frac{A}{\omega}\right)J_{k-n}\left(\frac{A}{\omega}\right)\left(1\!-\!\frac{1}{\epsilon_1\!+\!\epsilon_2\!+\!k\omega}\left(\frac{\epsilon_1\!\left(\epsilon_1\!+\!k\omega\right)}{\epsilon_1\!-\!g\!+\!n\omega}\!+\!\frac{\epsilon_2\!\left(\epsilon_2\!+\!k\omega\right)}{\epsilon_2\!-\!g\!+\!n\omega}\right)\right)\right]^2\right\}+\ldots.
\label{eq:average_probability_result}
\end{aligned}
\end{equation}
\end{widetext}
The analysis shows that the transition probability from the ground state to the highest excited level, $P_{1\to 4}$, is determined only in the second order of perturbation theory. From the obtained expressions \eqref{eq:average_probability_result}, it can be seen that the transition probabilities have a resonant character and the positions of multiphoton resonances are determined by the expressions:
\begin{subequations}\label{eq:res14}
\begin{align}
\left(\epsilon_{1,2}\pm g\right)/\omega\in\mathbb{Z},\label{eq:res14a}\\
\left(\epsilon_1+\epsilon_2\right)/\omega\in\mathbb{Z}.\label{eq:res14b}
\end{align}
\end{subequations}
As will be seen below from the direct numerical simulation, the largest transition probability $P_{1\to 4}$ is observed when the last of resonance conditions Eqs.~\eqref{eq:res14} is fulfilled, which does not depend on the coupling constant $g$.

Similar conditions for multiphoton resonance transitions $1\to2$ and $1\to3$ in the system of two coupled qubits were obtained in the rotating wave approximation \cite{Satanin2010,Satanin2012}, $\Delta_{q}\ll\omega$. In this case the resonance conditions depend on the coupling constant, $g$. However, the articles~\cite{Satanin2010,Satanin2012}, does not provide an analytical explanation for the appearance of stable multiphoton resonances with respect to the coupling parameter of qubits upon population inversion, and their study was carried out mainly on the basis of numerical simulations, in contrast to our current work.

\section{Rotating wave approximation of FLOQUET state}\label{sec:RWA}
In this section, we investigate the resonant dynamics, when at least one of the conditions $\left(\epsilon_{1,2}\pm g\right)/\omega\in\mathbb{Z}$ or $\left(\epsilon_{1}\pm\epsilon_{2}\right)/\omega\in\mathbb{Z}$ is satisfied. If the resonance conditions are satisfied, the theory developed in Section \ref{sec:floquet} becomes inapplicable due to the appearance of singular terms. The theory developed here allows to eliminate divergences in Eq.~\eqref{eq:average_probability_result} and determine the shape of the resonant peak. 

First we switch to the interaction representation with respect to the unperturbed Hamiltonian \eqref{eq:hamiltonian_unperturbed_part} for which the corresponding evolution operator $\hat{U}_0\left(t\right)$ is as follows:
\begin{align}
\hat{U}&_0\left(t\right)\!=\!\exp\left\{-i\:\mathrm{diag}\left[\gamma_1^{\left(0\right)}t\!-\!\frac{A}{\omega}\left(\sin\left(\omega t\!-\!\varphi_0\right)\!+\!\sin\varphi_0\right),\right.\right.\nonumber\\ 
&\left.\left.\gamma_2^{\left(0\right)}t,\gamma_3^{\left(0\right)}t,\gamma_4^{\left(0\right)}t\!+\!\frac{A}{\omega}\left(\sin\left(\omega t\!-\!\varphi_0\right)\!+\!\sin\varphi_0\right)\right]\right\}.\label{eq:evolution_operator_unperturbed}
\end{align}
The interaction Hamiltonian $\hat{H}_{1,I}\left(t\right)$ can then be shown to be
\begin{equation}
\begin{gathered}
\hat{H}_{1,I}\left(t\right)=-\frac{1}{2}\begin{pmatrix}
0 & \Delta_2\xi_2^{+*} & \Delta_1\xi_1^{+*} & 0\\
\Delta_2\xi_2^{+} & 0 & 0 & \Delta_1\xi_1^{-*}\\
\Delta_1\xi_1^{+} & 0 & 0 & \Delta_2\xi_2^{-*}\\
0 & \Delta_1\xi_1^{-} & \Delta_2\xi_2^{-} & 0
\end{pmatrix},
\end{gathered}
\label{eq:hamiltonian_interaction_representation}
\end{equation}
where the functions $\xi_q^{\pm}\left(t\right)$ are defined by
\begin{equation}
\begin{gathered}
\xi_q^{\pm}\!=\!\exp\!\left[i\!\left(\epsilon_q\pm g\right)\!t+i\frac{A}{\omega}\!\left(\sin\left(\omega t-\varphi_0\right)+\sin\varphi_0\right)\right]\!.
\end{gathered}
\label{eq:xi_q_pm_t}
\end{equation}
First consider the case when the only one condition $\delta_{2}^{+}=\epsilon_{2}+g+K_{2}^{+}\omega\approx 0$ is satisfied. This is the point at which the rotating wave approximation (RWA) is made. Using the generating function of the Bessel functions of the first kind, $e^{iz\sin q}=\sum_k J_k\left(z\right)e^{ikq}$, we can isolate slow oscillations of $\xi_{2}^{+}\left(t\right)$: $\xi_{2}^{+}\left(t\right)\approx J_{K_{2}^{+}}\left(A/\omega\right)\exp\left[i\left(\delta_{2}^{+}t+\frac{A}{\omega}\sin\varphi_0-K_{2}^{+}\varphi_0\right)\right]$. Similarly, oscillations of $\xi_{1}^{+}\left(t\right)$, $\xi_{1}^{-}\left(t\right)$ and $\xi_{2}^{-}\left(t\right)$ will quickly average to $0$ on any appreciable time scale, i.e., $\xi_{1}^{+}\approx\xi_{1}^{-}\approx\xi_{2}^{-}\approx 0$. Thus the Hamiltonian $\hat{H}_{1,I}\left(t\right)$ can be approximated in the interaction picture as
\begin{widetext}
\begin{equation}
\begin{gathered}
\hat{H}_{1,I}^{\left(2\right)}\left(t\right)=-\frac{\Delta_2}{2}J_{K_{2}^{+}}\left(\frac{A}{\omega}\right)\begin{pmatrix}
0 & e^{-i\left(\delta_{2}^{+}t+\frac{A}{\omega}\sin\varphi_0-K_{2}^{+}\varphi_0\right)} & 0 & 0\\
e^{i\left(\delta_{2}^{+}t+\frac{A}{\omega}\sin\varphi_0-K_{2}^{+}\varphi_0\right)} & 0 & 0 & 0\\
0 & 0 & 0 & 0\\
0 & 0 & 0 & 0
\end{pmatrix}.
\end{gathered}
\label{eq:hamiltonian_interaction_representation_rwa_2}
\end{equation}
\end{widetext}
The transition probability from state $\left|a\right>$ to state $\left|b\right>$, $P_{1\to 2}\left(t,t_0\right)$, calculated up to the lowest order in smallness parameters $\Delta_{q}$, is determined by the evolution operator element $\left(\hat{U}_{1,I}^{\left(2\right)}\left(t\right)\right)_{12}$ obeying the Cauchy problem
\begin{align}
\partial_{t}^{2}\!&\left(\hat{U}_{1,I}^{\left(2\right)}\right)_{21}\!-i\delta_2^{+}\partial_{t}\!\left(\hat{U}_{1,I}^{\left(2\right)}\right)_{21}\!+\left(\Omega_0^{\left(2\right)}\right)^2\!\left(\hat{U}_{1,I}^{\left(2\right)}\right)_{21}\!=0,\nonumber\\
&\left(\hat{U}_{1,I}^{\left(2\right)}\left(0\right)\right)_{21}\!=0,\label{eq:evolution_operator_2_21_cauchy_problem}\\
\partial_{t}\!&\left(\hat{U}_{1,I}^{\left(2\right)}\left(0\right)\right)_{21}\!=-i\left(\hat{H}_{1,I}^{\left(2\right)}\left(0\right)\right)_{21},\nonumber
\end{align}
where $\Omega_0^{\left(2\right)}=\frac{\Delta_2}{2}J_{K_2^{+}}\left(A/\omega\right)$. The solution is
\begin{equation}
\begin{gathered}
\left(\hat{U}_{1,I}^{\left(2\right)}\right)_{21}=i\frac{\Omega_0^{\left(2\right)}}{\Omega^{\left(2\right)}}e^{i\left(\frac{\delta_{2}^{+}}{2}t+\frac{A}{\omega}\sin\varphi_0-K_{2}^{+}\varphi_0\right)}\sin\Omega^{\left(2\right)} t,
\end{gathered}
\label{eq:evolution_operator_2_21}
\end{equation}
with frequency $\Omega^{\left(2\right)}=\sqrt{\left(\Omega_0^{\left(2\right)}\right)^2+\left(\delta_{2}^{+}/2\right)^2}$. Using the definition Eq.~\eqref{eq:average_probability_definition}, after averaging over the pulse length $\tau$ and initial phase $\varphi_0$, finally obtain
\begin{equation}
\begin{gathered}
\bar{P}_{1\to 2}=\dfrac{1}{2}\left[1+\left(\dfrac{\delta_{2}^{+}}{2\Omega_0^{\left(2\right)}}\right)^2\right]^{-1}.
\end{gathered}
\label{eq:average_probability_result_12}
\end{equation}
The average transition probability $\bar{P}_{1\to 2}$ curve has a half width at half maximum $\text{HWHM}=2\left|\Omega_0^{\left(2\right)}\right|\sim\Delta_2$.

In an entirely similar manner to that described earlier, the case if the only condition $\delta_{1}^{+}=\epsilon_{1}+g+K_{1}^{+}\omega\approx 0$ is satisfied can be studied using RWA approximation. So,
\begin{equation}
\begin{gathered}
\bar{P}_{1\to 3}=\dfrac{1}{2}\left[1+\left(\dfrac{\delta_{1}^{+}}{2\Omega_0^{\left(3\right)}}\right)^2\right]^{-1},
\end{gathered}
\label{eq:average_probability_result_13}
\end{equation}
where $\Omega_0^{\left(3\right)}=\frac{\Delta_1}{2}J_{K_1^{+}}\left(A/\omega\right)$. The curve half width at half maximum is $\text{HWHM}=2\left|\Omega_0^{\left(3\right)}\right|\sim\Delta_1$.

The last case that we will consider is the fulfillment of the only condition $\delta_{12}^{+}=\epsilon_{1}+\epsilon_{2}+K_{12}^{+}\omega\approx 0$. It is clear that the transition probability $\bar{P}_{1\to 4}\left(t,t_0\right)$ cannot be calculated using Hamiltonian \eqref{eq:hamiltonian_interaction_representation} within the RWA framework. To be able to go further, we apply additional unitary transformation to Eq.~$\eqref{eq:hamiltonian_interaction_representation}$ with a unitary operator $\hat{U}_{1,I}(t)=\exp\left(-i\int_{0}^{t}\text{d}\tau\hat{H}_{1,I}\left(\tau\right)\right)$.
Under this change, up to second order of smallness parameters $\Delta_{q}$ the Hamiltonian $\hat{H}_{1,I}\left(t\right)$ transforms into $\hat{H}_{2,I}\left(t\right)=\frac{i}{2}\int_{0}^{t}\text{d}\tau\left[\hat{H}_{1,I}\left(\tau\right),\hat{H}_{1,I}\left(t\right)\right]$. Nonzero Hamiltonian $\hat{H}_{2,I}\left(t\right)$  matrix elements have the following form:
\begin{align}
&\left(\hat{H}_{2,I}\left(t\right)\right)_{11}=
-\frac{1}{4}\Im\left(\Delta_1^2\xi_1^{+}\Xi_1^{+*}+\Delta_2^2\xi_2^{+}\Xi_2^{+*}\right),\nonumber\\
&\left(\hat{H}_{2,I}\left(t\right)\right)_{22}=
-\frac{1}{4}\Im\left(\Delta_1^2\xi_1^{-}\Xi_1^{-*}-\Delta_2^2\xi_2^{+}\Xi_2^{+*}\right),\nonumber\\
&\left(\hat{H}_{2,I}\left(t\right)\right)_{33}=
\frac{1}{4}\Im\left(\Delta_1^2\xi_1^{+}\Xi_1^{+*}-\Delta_2^2\xi_2^{-}\Xi_2^{-*}\right),\nonumber\\
&\left(\hat{H}_{2,I}\left(t\right)\right)_{44}=
\frac{1}{4}\Im\left(\Delta_1^2\xi_1^{-}\Xi_1^{-*}+\Delta_2^2\xi_2^{-}\Xi_2^{-*}\right),\nonumber\\
&\left(\hat{H}_{2,I}\left(t\right)\right)_{14}=
\frac{i}{8}\Delta_1\Delta_2\left(\xi_1^{-}\Xi_2^{+}-\xi_1^{+}\Xi_2^{-}\right.\label{eq:hamiltonian_interaction_representation_2_elements}\\
&\left.\hspace{4cm}+\xi_2^{-}\Xi_1^{+}-\xi_2^{+}\Xi_1^{-}\right)^{*},\nonumber\\
&\left(\hat{H}_{2,I}\left(t\right)\right)_{23}=
\frac{i}{8}\Delta_1\Delta_2\left(\xi_1^{+*}\Xi_2^{+}-\xi_1^{-*}\Xi_2^{-}\right.\nonumber\\
&\left.\hspace{4cm}+\xi_2^{-}\Xi_1^{-*}-\xi_2^{+}\Xi_1^{+*}\right),\nonumber\\
&\left(\hat{H}_{2,I}\left(t\right)\right)_{41}=\left(\hat{H}_{2,I}\left(t\right)\right)_{14}^{*},\nonumber\\
&\left(\hat{H}_{2,I}\left(t\right)\right)_{32}=\left(\hat{H}_{2,I}\left(t\right)\right)_{23}^{*},\nonumber
\end{align}
where the functions $\Xi_q^{\pm}\left(t\right)$ are defined by
\begin{equation}
\begin{gathered}
\Xi_q^{\pm}\left(t\right)=\int_{0}^{t}\text{d}\tau\xi_q^{\pm}\left(\tau\right).
\end{gathered}
\label{eq:Xi_q_pm_t}
\end{equation}
Nonzero matrix elements of a new Hamiltonian $\hat{H}_{2,I}^{\left(4\right)}\left(t\right)$ obtained from Eq.~\eqref{eq:hamiltonian_interaction_representation_2_elements} by applying RWA look like 
\begin{widetext}
\begin{gather}
\left(\hat{H}_{2,I}^{\left(4\right)}\left(t\right)\right)_{11}=
-\frac{1}{4}\sum_{k=-\infty}^{+\infty}J_k^2\left(\frac{A}{\omega}\right)\left(\frac{\Delta_1^2}{\epsilon_{1}+g+k\omega}+\frac{\Delta_2^2}{\epsilon_{2}+g+k\omega}\right),\nonumber\\
\left(\hat{H}_{2,I}^{\left(4\right)}\left(t\right)\right)_{22}=
-\frac{1}{4}\sum_{k=-\infty}^{+\infty}J_k^2\left(\frac{A}{\omega}\right)\left(\frac{\Delta_1^2}{\epsilon_{1}-g+k\omega}-\frac{\Delta_2^2}{\epsilon_{2}+g+k\omega}\right),\nonumber\\
\left(\hat{H}_{2,I}^{\left(4\right)}\left(t\right)\right)_{33}=
\frac{1}{4}\sum_{k=-\infty}^{+\infty}J_k^2\left(\frac{A}{\omega}\right)\left(\frac{\Delta_1^2}{\epsilon_{1}+g+k\omega}-\frac{\Delta_2^2}{\epsilon_{2}-g+k\omega}\right),\\
\left(\hat{H}_{2,I}^{\left(4\right)}\left(t\right)\right)_{44}=
\frac{1}{4}\sum_{k=-\infty}^{+\infty}J_k^2\left(\frac{A}{\omega}\right)\left(\frac{\Delta_1^2}{\epsilon_{1}-g+k\omega}+\frac{\Delta_2^2}{\epsilon_{2}-g+k\omega}\right),\nonumber\\
\left(\hat{H}_{2,I}^{\left(4\right)}\left(t\right)\right)_{14}=
\Delta_1\Delta_2\frac{g}{4}e^{-i\left(\delta_{12}^{+}t+2\frac{A}{\omega}\sin\varphi_0-K_{12}^{+}\varphi_0\right)}\sum_{k=-\infty}^{+\infty}J_k\left(\frac{A}{\omega}\right)J_{K_{12}^{+}-k}\left(\frac{A}{\omega}\right)\left(\frac{1}{\left(\epsilon_{1}+k\omega\right)^2-g^2}+\frac{1}{\left(\epsilon_{2}+k\omega\right)^2-g^2}\right).\nonumber
\label{eq:hamiltonian_interaction_representation_rwa_4_elements}
\end{gather}
\end{widetext}
In the lowest order in smallness parameters the transition probability $P_{1\to 4}\left(t,t_0\right)$ is determined by the evolution operator element $\left(\hat{U}_{1,I}^{\left(4\right)}\left(t\right)\right)_{14}$ obeying the Cauchy problem
\begin{align}
\partial_{t}^{2}\!&\left(\hat{U}_{2,I}^{\left(4\right)}\right)_{41}\!-i\left[\delta_{12}^{+}+\delta_0-2\left(\hat{H}_{2,I}^{\left(4\right)}\right)_{11}\right]\partial_{t}\!\left(\hat{U}_{2,I}^{\left(4\right)}\right)_{41}\nonumber\\
&+\left[\left(\Omega_0^{\left(4\right)}\right)^2\!-\left(\delta_{12}^{+}-\left(\hat{H}_{2,I}^{\left(4\right)}\right)_{11}\right)\right.\nonumber\\
&\left.\hspace{1.5cm}\times\left(\delta_{0}-\left(\hat{H}_{2,I}^{\left(4\right)}\right)_{11}\right)\right]\left(\hat{U}_{2,I}^{\left(4\right)}\right)_{41}=0,\nonumber\\
&\left(\hat{U}_{2,I}^{\left(4\right)}\left(0\right)\right)_{41}\!=0,\label{eq:evolution_operator_4_41_cauchy_problem}\\
\partial_{t}\!&\left(\hat{U}_{2,I}^{\left(4\right)}\left(0\right)\right)_{41}\!=-i\left(\hat{H}_{2,I}^{\left(4\right)}\left(0\right)\right)_{41},\nonumber
\end{align}
where
\begin{align}
&\delta_{0}\!=\!-\frac{1}{2}\!\sum_{k=-\infty}^{+\infty}\!J_k^2\!\left(\frac{A}{\omega}\right)\!\left(\!\frac{\Delta_1^2(\epsilon_{1}\!+\!k\omega)}{\left(\epsilon_{1}\!+\!k\omega\right)^2\!-\!g^2}\!+\!\frac{\Delta_2^2(\epsilon_{2}\!+\!k\omega)}{\left(\epsilon_{2}\!+\!k\omega\right)^2\!-\!g^2}\!\right)\!,\nonumber\\
&\Omega_0^{\left(4\right)}\!=\!\frac{g\Delta_1\Delta_2}{4}\sum_{k=-\infty}^{+\infty}\!J_k\!\left(\frac{A}{\omega}\right)\!J_{K_{12}^{+}-k}\left(\frac{A}{\omega}\right)\label{eq:delta_0_Omega_0_4}\\
&\hspace{2.cm}\times\left(\frac{1}{\left(\epsilon_{1}\!+\!k\omega\right)^2\!-\!g^2}+\frac{1}{\left(\epsilon_{2}\!+\!k\omega\right)^2-\!g^2}\right)\!.\nonumber
\end{align}
Solving the problem Eq.~\eqref{eq:evolution_operator_4_41_cauchy_problem}, we find
\begin{align}
\left(\hat{U}_{2,I}^{\left(4\right)}\left(t\right)\right)_{41}\!&=-i\frac{\Omega_0^{\left(4\right)}}{\Omega^{\left(4\right)}}\exp\!\left[i\left(\!\left(\frac{\delta_{12}^{+}+\delta_{0}}{2}\!-\!\left(\hat{H}_{2,I}^{\left(4\right)}\right)_{11}\right)\!t\right.\right.\nonumber\\
&\left.\left.+2\frac{A}{\omega}\sin\varphi_0\!-\!K_{12}^{+}\varphi_0\right)\right]\sin\Omega^{\left(4\right)}t,\label{eq:evolution_operator_2_41}
\end{align}
where frequency
$\Omega^{\left(4\right)}=\sqrt{\left(\Omega_0^{\left(4\right)}\right)^2+\left(\frac{\delta_{12}^{+}-\delta_{0}}{2}\right)^2}$. Finally,
\begin{equation}
\begin{gathered}
\bar{P}_{1\to 4}=\dfrac{1}{2}\left[1+\left(\dfrac{\delta_{12}^{+}-\delta_{0}}{2\Omega_0^{\left(4\right)}}\right)^2\right]^{-1}.
\end{gathered}
\label{eq:average_probability_result_14}
\end{equation}
The corresponding curve $\text{HWHM}=2\left|\Omega_0^{\left(4\right)}\right|\sim\Delta_1\Delta_2$.

\section{Numerical results and discussion}

In this section the multiphoton resonances formation is discussed in detail based on the developed analytical approach and numerical analysis within the Floquet theory framework. We focus on studying the qubits response in a monochromatic field of a fixed non-resonant frequency when the control bias parameters $\epsilon_q$ and the coupling strength $g$ are changed in a wide scanning range, similar to the implemented experiments \cite{Izmalkov2008,Ilichev2010}.

\subsection{Multiphoton interference effects}

\begin{figure}\center
	\includegraphics[width=0.97\columnwidth]{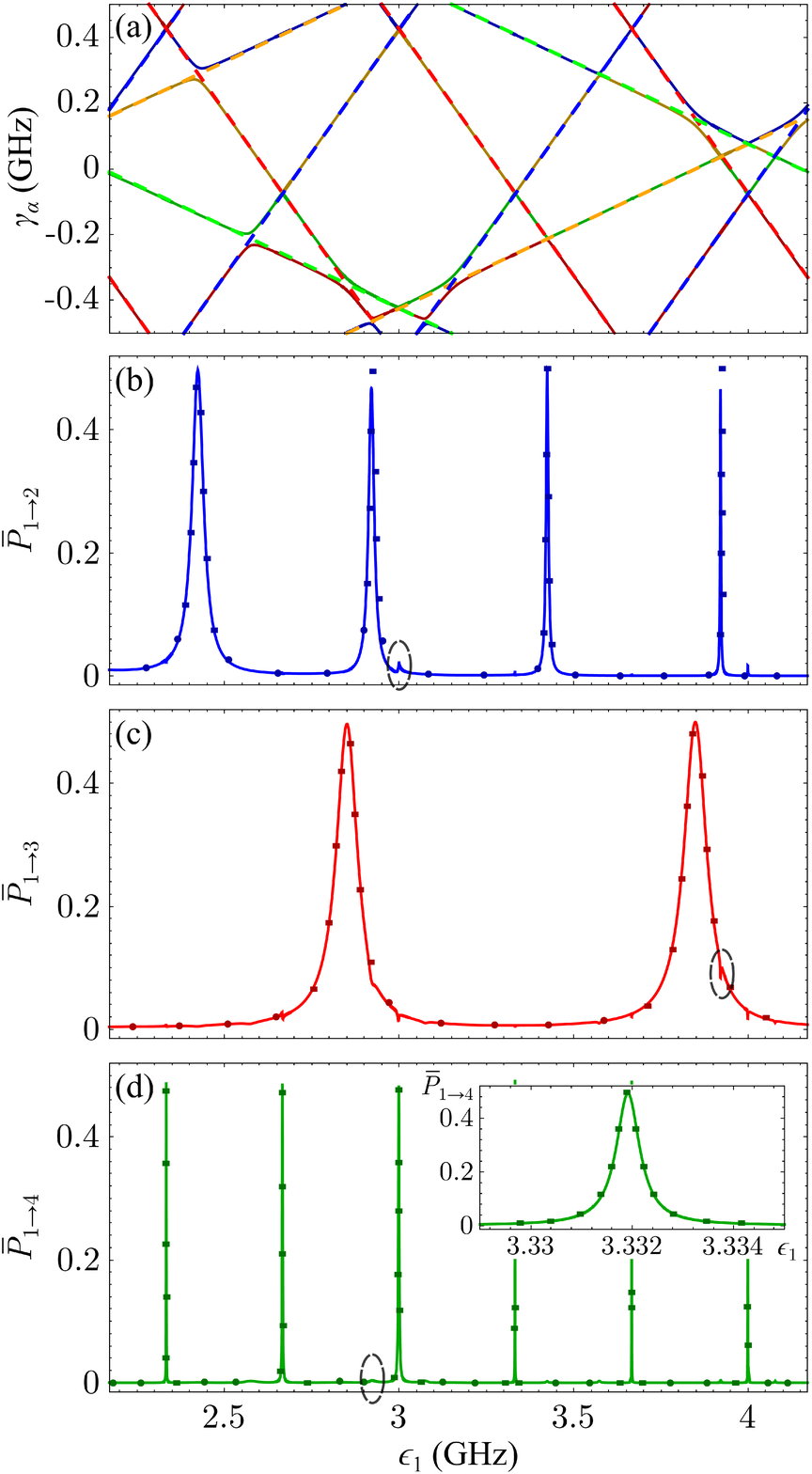}
\caption{(a) Quasienergies $\gamma_{\alpha}$ and (b,c,d) average transition probabilities $\bar{P}_{1\to b}$ ($b = 2, 3, 4$) as functions of the control parameter $\epsilon_1$. The results of direct numerical calculations (solid curves) compared with analytical results: round markers are derived from non-resonant expressions~\eqref{eq:average_probability_result}; square ones are derived from resonant expressions~\eqref{eq:average_probability_result_12}, \eqref{eq:average_probability_result_13} and \eqref{eq:average_probability_result_14}. The following parameters were used: $\Delta_1=0.1$~GHz, $\Delta_2=0.15$~GHz, $g=0.15$~GHz, $A=5$~GHz, $\omega=1$~GHz, $\epsilon_2=2\epsilon_1$.}
\label{fig:fig1}
\end{figure}

As is known, the system of two coupled qubits in a time-dependent ﬁeld of arbitrary amplitude has four quasienergy levels $\gamma_{\alpha}$, which undergo anticrossing when the driving external fields change (see Fig.~\ref{fig:fig1}~(a)), affecting the multiphoton transitions formation and the entanglement generation \cite{Sauer2012,Quintana2013,Gramajo2017,Gramajo2018}.

\begin{figure}[b]\center
	\includegraphics[width=0.95\columnwidth]{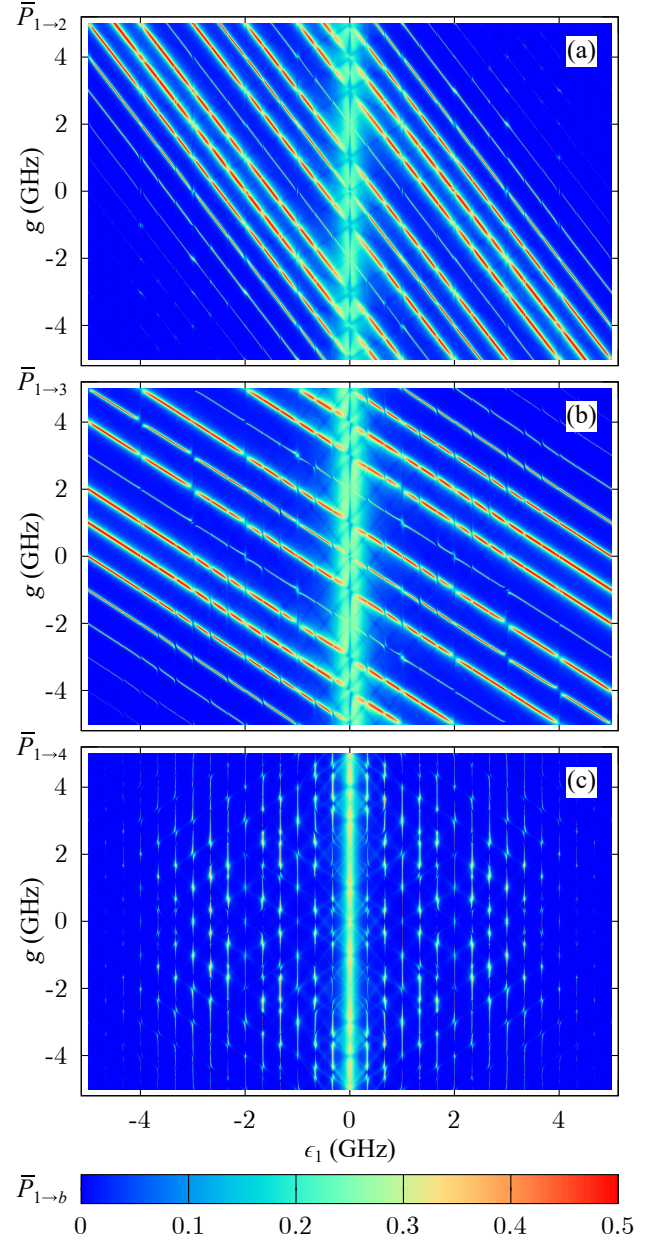}
\caption{Transition probabilities $\bar{P}_{1\to b}$ of two qubits as functions of the control parameter $\epsilon_1$ and the coupling parameter $g$. Parameters: $\Delta_1=0.2$~GHz, $\Delta_2=0.3$~GHz, $A=5$~GHz, $\omega=1$~GHz, $\epsilon_2=2\epsilon_1$.}
\label{fig:fig2}
\end{figure} 

Fig.~\ref{fig:fig1} shows the dependencies of the coupled qubits quasienergies and multiphoton transition probabilities as a function of energy bias $\epsilon_1$ for the case $s=\epsilon_2/\epsilon_1=2$. According to the solution of Eq.~\eqref{eq:equation_floquet}, the dependence of quasienergies is shown in Fig.~\ref{fig:fig1}~(a). It is known that the behavior of quasienergies depends on the symmetry class: if they belong to the same symmetry class, then they cross; otherwise they form an anticrossing (that corresponds to the diabatic states and the adiabatic states). Let us note that each of the quasienergy crossing can be associated with the regions of multiphoton transitions. So each crossing of the red and orange dashed curves on Fig.~\ref{fig:fig1}~(a) corresponds to the main transition peak $1\to2$, the crossing of the red and green curves corresponds to the main peak $1\to3$, and the crossing of the red and blue curves corresponds to the inversion transition $1\to4$.  It is seen that the shapes and positions of the resonance curves obtained by us numerically Eq.~\eqref{eq:average_probability} in good agreement with the results of the expressions found in the framework of the Floquet perturbation theory: analytical dependencies for non-resonant Eq.~\eqref{eq:average_probability_result} and resonant cases Eqs.~\eqref{eq:average_probability_result_12}, \eqref{eq:average_probability_result_13}, \eqref{eq:average_probability_result_14} are plotted with round and square markers respectively. An important property is the resonance peaks $\bar{P}_{1\to 3}\left(\epsilon_1\right)$ shift by the coupling parameter $g$ value (for the transition $\bar{P}_{1\to 2}\left(\epsilon_1\right)$ by $g/s$). From this shift magnitude, one can experimentally determine both the coupling type (the left shift corresponds to ferromagnetic coupling, shown in Fig.~\ref{fig:fig1}~(b, c), and the right shift to antiferromagnetic one), and the coupling strength $g$ value. Similar reasoning can be carried out for the case of scanning over the second qubit energy bias $\epsilon_2$.
\begin{figure*}[htb]\center
	\includegraphics[width=0.95\linewidth]{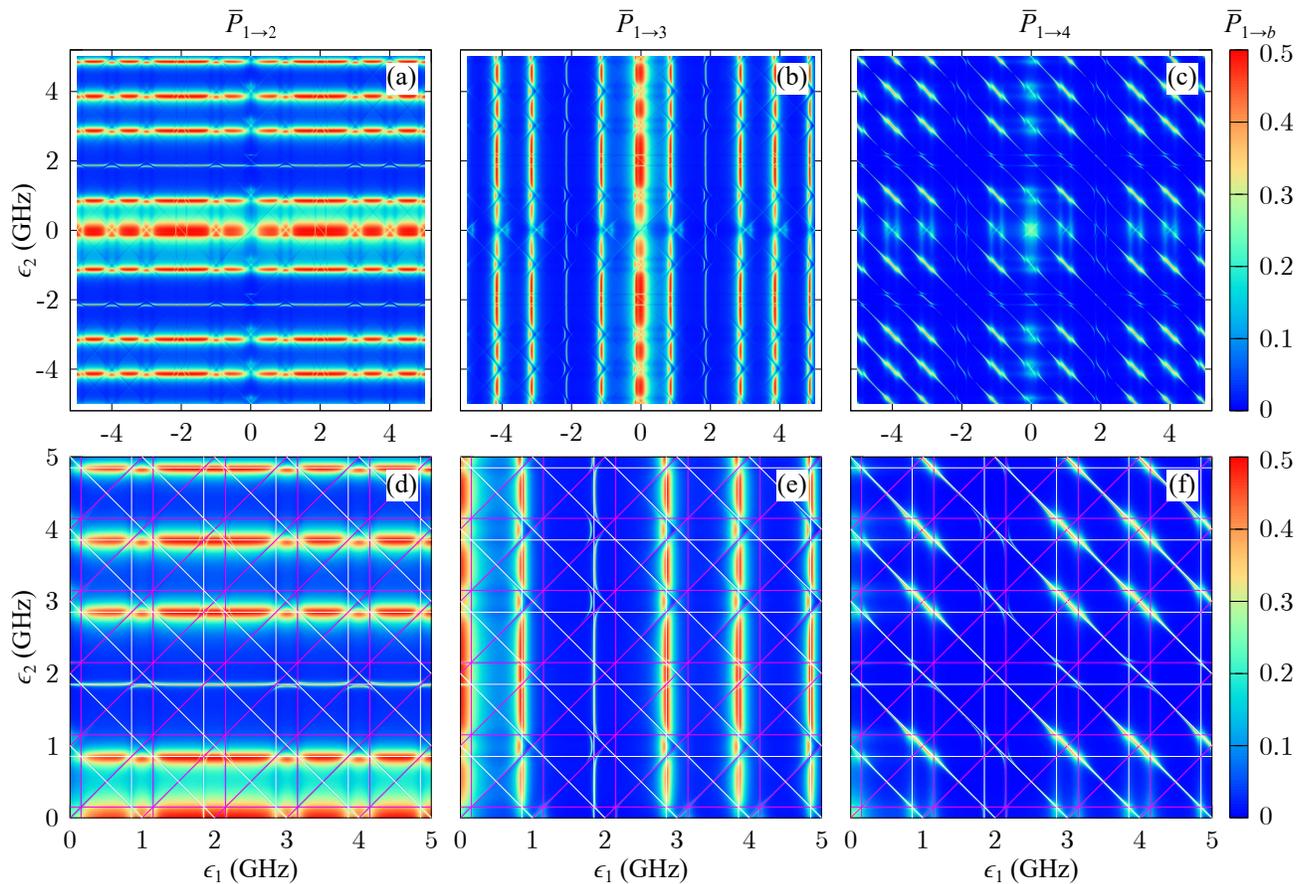}\\
\caption{Transition probabilities $\bar{P}_{1\to 2}$ (a, d), $\bar{P}_{1\to 3}$ (b, e), and $\bar{P}_{1\to 4}$ (c, f) of two coupled qubits as functions of the control parameters $\epsilon_1$ and $\epsilon_2$. The plots (d, e, f) show the enlarged areas from (a, b, c), respectively; horizontal and vertical lines correspond to resonance condition Eqs.~\eqref{eq:res14a}, diagonal ones to Eqs.~\eqref{eq:res14b}. Parameters: $\Delta_1=0.2$~GHz, $\Delta_2=0.3$~GHz, $g=0.15$~GHz, $A=5$~GHz, $\omega=1$~GHz.}
\label{fig:fig3}
\end{figure*}

Multiphoton resonance transition from the ground state to the highest excited level have a different character. It can be seen from Fig.~\ref{fig:fig1}~(d) that the main resonance curves $\bar{P}_{1\to 4}(\epsilon_1)$ has not dependent of coupling $g$ (see also  Fig.~\ref{fig:fig2}~(c)) and their positions are determined by Eq.~\eqref{eq:average_probability_result}. The main peaks positions obey Eq.~\eqref{eq:res14b}; their formation was discussed on the resonant perturbation theory in Sec.~\ref{sec:RWA}. The Fig.~\ref{fig:fig1}~(d) inset demonstrates a good agreement between theoretical and numerical results. The positions of low intensity side peaks correspond to Eqs.~\eqref{eq:res14a}. They can be characterized as interference peaks due to transitions through intermediate levels. This fact can clearly be understood from the analysis in expression~\eqref{eq:average_probability_result}: the first term in Eq.~\eqref{eq:average_probability_result} corresponds to the simultaneous fulfillment of two resonant conditions $\left(\epsilon_{2} + g\right)/\omega\in\mathbb{Z}$ and $\left(g - \epsilon_{1}\right)/\omega\in\mathbb{Z}$ for the transition $1\to2$ and $2\to4$, respectively. The second term in Eq.~\eqref{eq:average_probability_result} corresponds to the fulfillment of the conditions  $\left(\epsilon_{1} + g\right)/\omega\in\mathbb{Z}$ and $\left(g - \epsilon_{2}\right)/\omega\in\mathbb{Z}$ for the transition $1\to3$ and $3\to4$ (see the expression \eqref{eq:S_matrix_elements} in the Appendix).

The analysis performed in Sec.~\ref{sec:RWA} showed that the resonance peaks of interlevel transitions in interacting two qubit system have a Lorentzian shape at $\Delta_{1,2}\ll A,\epsilon_{1,2},g$. According to the obtained expressions~\eqref{eq:average_probability_result_12}, \eqref{eq:average_probability_result_13} and~\eqref{eq:average_probability_result_14} the main peaks reach the maximum value $0.5$, similarly to the two-level system case \cite{Shirley1965}. It is shown that the peak widths depend on the qubits tunneling energies $\Delta_{1,2}$ (the corresponding HWHM of $\bar{P}_{1\to 2}$, $\bar{P}_{1\to 3}$, $\bar{P}_{1\to 4}$ are $\sim\Delta_2$, $\sim\Delta_1$, $\sim\Delta_1\Delta_2$, respectively) and are fully consistent with the numerical simulations. The quantum-coherent tunneling is violated in the region between resonances and the transition probabilities $\bar{P}_{1\to 2}$, $\bar{P}_{1\to 3}$ are highly dependent on the non-resonant background determined by the driving field amplitude $A$ included in the Bessel function arguments, $\sim J_n\left(A/\omega\right)$. This statement follows from the Floquet resonance perturbation theory, which allows one to determine the background level at small $\Delta_{1,2}$ by the expressions for $\Omega_0^{\left(2,3\right)}$. The peaks become narrow for high-order multiphoton resonances ($n\gg 1$), which is clearly seen in Fig.~\ref{fig:fig1}~(a). The non-resonant background influence on the transition $1\to 4$ is less pronounced since $\Omega_0^{\left(4\right)}$ value is expressed through the products of the Bessel functions, which significantly weakens the contribution to the background.

In a number of experiments the coupling strength between qubits can be controlled \textit{in situ} \cite{Plourde2004,Ploeg2007,Groszkowski2011,Allman2014}. The multiphoton transition probabilities maps with a simultaneous change in the control bias $\epsilon_1$ and $g$ are shown in the Fig.~\ref{fig:fig2}.

A different kind of excitation is seen here: the bright slant multiphoton transition lines for jumps to intermediate nonlocal states $1\to 2$ (Fig.~\ref{fig:fig2}~(a)) and $1\to 3$ (Fig.~\ref{fig:fig2}~(b)) are observed, which correspond to the parameters when qubits are not entangled. The inclinations of the multiphoton resonance lines can be used to establish the bias parameters ratio of individual qubits. There is a grid of main and fractional resonances insensitive to the coupling strength $g$ for $1\to 4$ transition. The obtained results are in good agreement with the data on the control and entanglement generation in the qubit systems~\cite{Gramajo2017}.

The multiphoton resonance interference patterns with a mutual change in the qubits bias parameters, similar to \cite{Izmalkov2008,Ilichev2010} experiments, were analyzed. Bright regions of the Landau--Zener tunneling in Fig.~\ref{fig:fig3} correspond to the obtained resonance conditions.

The resonance lines are plotted in Fig.~\ref{fig:fig3}~(d, e, f) in order to understand the numerical results. Horizontal and vertical lines correspond to condition~\eqref{eq:res14a}, diagonal ones to~\eqref{eq:res14b}, namely
\begin{align*}
&\left(\epsilon_1+g\right)/\omega\in\mathbb{Z},&&\text{white vertical lines},\\
&\left(\epsilon_1-g\right)/\omega\in\mathbb{Z},&&\text{purple vertical lines},\\
&\left(\epsilon_2+g\right)/\omega\in\mathbb{Z},&&\text{white horizonal lines},\\
&\left(\epsilon_2-g\right)/\omega\in\mathbb{Z},&&\text{purple horizonal lines},\\
&\left(\epsilon_1+\epsilon_2\right)/\omega\in\mathbb{Z},&&\text{white diagonal lines},\\
&\left(\epsilon_1-\epsilon_2\right)/\omega\in\mathbb{Z},&&\text{purple diagonal lines}.
\end{align*}
For example, the $1\to 3$ transition dominant resonance condition is $\left(\epsilon_1+g\right)/\omega\in\mathbb{Z}$, which is marked by white vertical lines $\sim\Delta_1$ wide. Probability features associated with interference effects due to cascade transitions through intermediate states are observed on these lines, for example, $1\to 2\to 3$ or $1\to 4\to 3$, which is confirmed by the resonance grid in Fig.~\ref{fig:fig3}~(d--f). These interference cascade transitions lead to the observation of low-intensity peaks (see markers in the  Fig.~\ref{fig:fig1}) and Fig.~\ref{fig:fig4}~(b--d).

\subsection{Dissipation effect on multiphoton resonances}
There are many dissipative processes under real experimental conditions affecting the results of measuring both individual and coupled qubits interference patterns. Following the ideology of the articles~\cite{Grifoni1998,Gramajo2018}, we describe dissipation in a two-qubit system under the assumption that each qubit interacts with its own bosonic reservoir. The noise spectrum is considered to be smooth that allows to use the Born-Markov approximation when studying the system dynamics~\cite{Kohler1997,Hone2009}. We get the Floquet-Markov master equation for two coupled qubits:
\begin{equation}
\frac{\partial\hat{\rho}}{\partial t}=-i\left[\hat{H}\left(t\right),\hat{\rho}\right]+\hat{\Gamma}\hat{\rho},
\label{eq:master_equation}
\end{equation}
with dissipative operator
\begin{equation}
\hat{\Gamma}=\sum_{q=1}^{2}\!\left(\Gamma_{\varphi_q}\hat{D}\!\left[\hat{\sigma}_{z}^{\left(q\right)}\right]+\Gamma_q\hat{D}\!\left[\hat{\sigma}_{-}^{\left(q\right)}\right]+\Gamma^\prime_{q}\hat{D}\!\left[\hat{\sigma}_{+}^{\left(q\right)}\right]\right)\!,
\end{equation}
where $\Gamma_{\varphi_q}$, $\Gamma_q$ and $\Gamma^\prime_q$ are dephasing, relaxation and excitation rates, respectively;  $\hat{D}\left[\hat{a}\right]\hat{\rho}\equiv\hat{a}\hat{\rho}\hat{a}^{\dag}-\frac{1}{2}\left\{\hat{a}^{\dag}\hat{a},\hat{\rho}\right\}$ and the Lindblad operators $\hat{\sigma}_z^{\left(q\right)}$, $\hat{\sigma}_{+}^{\left(q\right)}$, $\hat{\sigma}_{-}^{\left(q\right)}$ are expressed via individual qubits (as if there were no coupling, $g=0$, and field, $v_{q}\left(t\right) = 0$) eigenstates $\left|\downarrow^{\left(q\right)}\right\rangle$, $\left|\uparrow^{\left(q\right)}\right\rangle$ as follows
\begin{equation}
\begin{aligned}
&\hat{\sigma}_z^{\left(q\right)}=\left|\uparrow^{\left(q\right)}\right\rangle\left\langle\uparrow^{\left(q\right)}\right|-\left|\downarrow^{\left(q\right)}\right\rangle\left\langle\downarrow^{\left(q\right)}\right|,\\
&\hat{\sigma}_{+}^{\left(q\right)}=\left|\uparrow^{\left(q\right)}\right\rangle\left\langle\downarrow^{\left(q\right)}\right|,\\
&\hat{\sigma}_{-}^{\left(q\right)}=\left|\downarrow^{\left(q\right)}\right\rangle\left\langle\uparrow^{\left(q\right)}\right|.
\end{aligned}
\end{equation}
At the reservoirs fundamental temperature $\tau_B$, the relaxation and excitation parameters are related as $\Gamma^\prime_q=\Gamma_q\exp\!\left(-\Delta E^{\left(q\right)}/\tau_B\right)$, where $\Delta E^{\left(q\right)}$ is the energy gap of the $q$-th qubit. In the numerical simulations we also take into account that the typical driving pulse duration $\tau=1-10\mu s$ in the experimental Landau--Zener interference measurements for flux qubits \cite{Berns2006,Oliver2005} corresponds to the times of the dissipative processes as $1/\Gamma_q<\tau<1/\Gamma_{\varphi_q}$.

\begin{figure}\center
	\includegraphics[width=1.\columnwidth]{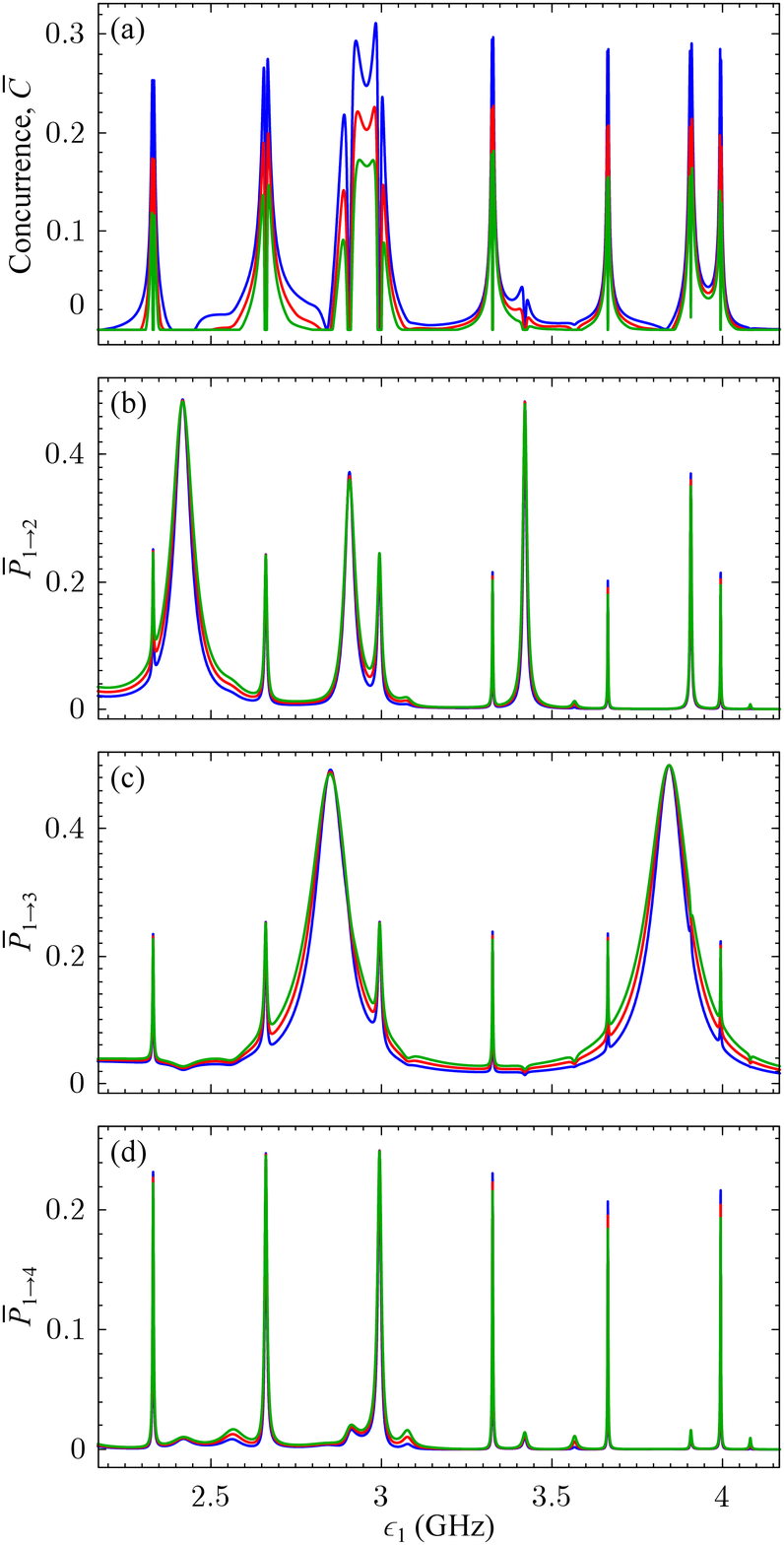}
\caption{(a) Average concurrence $\bar{C}$ and (b,c,d) average transition probabilities $\bar{P}_{1\to b}$ as functions of the control parameter $\epsilon_1$. Curves of various colors correspond to different dephasing rates: $\Gamma_{\varphi_1}=\Gamma_{\varphi_2}=0$ GHz (blue), $\Gamma_{\varphi_1}=\Gamma_{\varphi_2}=10^{-5}$ GHz (red), $\Gamma_{\varphi_1}=\Gamma_{\varphi_2}=2\cdot10^{-5}$ GHz (green). Reservoirs temperature is 30 mK. The parameters of qubits are the same as in Fig.~\ref{fig:fig1}.}
\label{fig:fig4}
\end{figure}

Using the Floquet theory and performing averaging over the initial phase $\varphi_0$ and the pulse length $\tau$, similar to Sec.~\ref{sec:floquet}, we can obtain that the average transition probabilities are determined as
\begin{equation}
\bar{P}_{\alpha\to\beta}=\frac{1}{T}\int\limits_0^T{\text{d}t\text{Tr}\left(\left|\beta\right\rangle\left\langle\beta\right|\hat\rho_T\left(t\right)\right)},
\end{equation}
where $\rho_T\left(t\right)$ is the periodic solution of~\eqref{eq:master_equation}, $\rho\left(t+T\right)=\rho\left(t\right)$, which for nonzero $\Gamma_q$ is unique. 

For a deeper understanding of interference processes when tuning the control field parameters, we analyze the entanglement of the states of coupled qubits. We calculate the average entanglement measure as a concurrence \cite{Wootters1998}: $\bar{C}=\max\left\{0,\lambda_4-\lambda_3-\lambda_2-\lambda_1\right\}$, where $\lambda_i$'s are real numbers in decreasing order and correspond to the eigenvalues of the matrix $R=\sqrt{\sqrt{\rho}\Tilde{\rho}\sqrt{\rho}}$, with $\Tilde{\rho}=\sigma^{\left(1\right)}_y\otimes\sigma^{\left(2\right)}_y\rho^{*}\sigma^{\left(1\right)}_y\otimes\sigma^{\left(2\right)}_y$. As can be seen from Fig.~\ref{fig:fig4}~(a), the entanglement generation occurs when the control bias parameter, $\epsilon_1$ , approaches the regions of multiphoton resonances of the inversion transition $1\to 4$ according to Eq.~\eqref{eq:res14b}. Note that if the main resonant peaks are satisfied for the $1\to2$ and $1\to3$ transitions Eqs.~\eqref{eq:res14a}, the entanglement in the system is destroyed. Thus, with an analytical explanation of the formation of the quantum tunneling suppression regions and multiphoton processes obtained in Sec.~\ref{sec:floquet}, we can implement controlled control and manipulation of entanglement even when taking into account the qubit coupling strength, $g$. This effect can be useful for measuring quantum state tomography \cite{Quintana2013, Roch2014}.

Our calculation in Fig.~\ref{fig:fig4} showed that the position of multiphoton resonances does not depend on dephasing effects. A similar statement is also true for the positions of antiresonances, in which quantum population trapping occurs. It is follows from the Floquet perturbation theory (see Sec.~\ref{sec:floquet}) that the width of resonant transitions to excited states (except transition to the highest level) is determined by the qubits tunneling energies: the smaller the parameters $\Delta_q$, the resonances are narrower. It follows from the numerical analysis in the framework of solving the equation for the density matrix that an increase in phase noise (dephasing rates $\Gamma_{\varphi_{q}}$) also affects the width of multiphoton resonances. As the dephasing rates increase, the peaks broaden and the intensity decreases, i.e. decreases the transition probability values (see Fig.~\ref{fig:fig4}). The overlapping of resonances occurs at $\gamma_{1,2}\approx\hbar\omega/2$, which was also observed for the single qubit case~\cite{Gelman2010,Satanin2014}. The strongest influence of noise is observed for the inverse population of the excited state with respect to the ground state $1\to 4$: corresponding intensity peaks are almost two times smaller than in the case of a system isolated from the external environment. This noise effect is due to the fact that the transition probability to a higher level is of the second order of smallness in tunneling constants. 

The effect of phase noise on the system is expressed in the interference patterns contrast and blurring loss (for example, Fig.~\ref{fig:fig3}) at higher dephasing rates. However, this does not prevent us from distinguishing the shift of resonances by the coupling constant $g$ value (for example, Fig.~\ref{fig:fig4} shows a shift of transition $1\to 3$ resonances in the dissipation presence). Therefore, we can conclude that the coupling constant $g$ can be determined experimentally within the framework of the two qubits amplitude spectroscopy method; the dephasing rate in the system can also be found from the interference pattern blurring degree. In addition, even under the noise effect it is possible to distinguish fractional resonances for the $1\to 4$ transition.

\section{Conclusion}
In this article, we expand Floquet's theory to study multiphoton processes in a system of two coupled superconducting flux qubits under the influence of a strong driving field. Numerical and analytical solutions in the framework of the generalized Floquet formalism are used to explain the multiphoton resonances and interference processes between qubits and the exciting alternating field. Within the framework of the perturbation theory in the Floquet basis, we obtained analytical solutions for the transition probabilities between the levels of coupled qubits in the computational basis up to the second order of smallness in the tunneling splitting of qubits. It is shown that this approach allows one to accurately predict the position of multiphoton resonances for an arbitrary amplitude of the external control field. The quasienergies and time-averaged transition probabilities are plotted, showing multiphoton resonance transitions occuring at anticrossing points. The developed theory allows to explain the complex interference pattern appearance with a simultaneous change in the parameters of the displacement of the qubits, which was previously observed in the~\cite{Izmalkov2008,Ilichev2010} experiments. It is shown that a specific feature of interlevel transitions ($1\to 2$, $1\to 3$) in the subspaces of individual qubits is the shift of the resonance peaks by the value of the coupling strength between the qubits. In this case, the width at half maximum of the resonances is determined by the qubit tunneling energy ($\sim\Delta_q$), and their intensity cannot exceed $0.5$, which follows from our detailed analysis in the RWA approximation near the found resonance conditions. A detailed analytical description for the inverse population of the levels $1\to 4$ formation is given. The main peaks of the direct multiphoton transitions are stable with respect to the displacements of the qubits and their coupling parameter, and their width is $\sim\Delta_1\Delta_2$. The formation of side peaks with low intensity occurs due to ladder transitions with the participation of intermediate levels when pairs of resonance conditions are fulfilled at once. In addition, cascade transitions significantly manifest themselves under the influence of the environment which was numerically studied using the Floquet--Markov equation. It is shown that dissipation in the system affects the width and intensity of the resonance peaks, but their position is unchanged. Consequently, the discovered interference effects can be successfully observed in experiments on amplitude spectroscopy \cite{Berns2006,Oliver2005,Rudner2008,Izmalkov2008,Ilichev2010} and the rate of phase relaxation in the system can be estimated from the degree of the pattern blurring. Our calculation also demonstrates the possibility to control entanglement in a system of a coupled qubits. When the system's displacement parameter is tuned to the region near the cascade transition, when multiphoton transitions effectively occur in both qubits, the concurrence of the system increases, and when it tuned to the main resonances $1\to 2$ and $1\to 3$, the entanglement in the system is completely suppressed. The general method described in our work provides a unified theoretical approach covering a wide range of the parameter space, and also compensates for the gap and indirect assumptions \cite{Ilichev2010,Satanin2012,Gramajo2018} when explaining the formation of multiphoton transition regions in a two-qubit system. The Floquet theory application to the study of superconducting qubits leads us to a better understanding of the spectroscopic measurement results and the dynamics of ac-controlled qubits, which are important for better characterizing and improving the performance of qubits.

\acknowledgments
The work was supported by the Competitiveness Improvement UNN program 5-100.

\appendixname{: PERTURBATION THEORY}

Considering tunnel splitting energies $\Delta_q$ is small enough and using them as a stationary perturbation theory parameters up to second order of smallness one can find eigenstates of the two uncoupled qubits system (first term in \eqref{eq:hamiltonian_global}):
\begin{equation}
\hspace*{-0.1cm}
\begin{aligned}
&\left|1\right>\!=\!\left|\downarrow\downarrow\right>\!=\!\left(\!1-\frac{\Delta_1^2}{8\epsilon_1^2}-\frac{\Delta_2^2}{8\epsilon_2^2},\frac{\Delta_2}{2\epsilon_2},\frac{\Delta_1}{2\epsilon_1},\frac{\Delta_1\Delta_2}{4\epsilon_1\epsilon_2}\!\right)^T\!,\\
&\left|2\right>\!=\!\left|\downarrow\uparrow\right>\!=\!\left(\!-\frac{\Delta_2}{2\epsilon_2},1-\frac{\Delta_1^2}{8\epsilon_1^2}-\frac{\Delta_2^2}{8\epsilon_2^2},-\frac{\Delta_1\Delta_2}{4\epsilon_1\epsilon_2},\frac{\Delta_1}{2\epsilon_1}\!\right)^T\!,\\
&\left|3\right>\!=\!\left|\uparrow\downarrow\right>\!=\!\left(\!-\frac{\Delta_1}{2\epsilon_1},-\frac{\Delta_1\Delta_2}{4\epsilon_1\epsilon_2},1-\frac{\Delta_1^2}{8\epsilon_1^2}-\frac{\Delta_2^2}{8\epsilon_2^2},\frac{\Delta_2}{2\epsilon_2}\!\right)^T\!,\\
&\left|4\right>\!=\!\left|\uparrow\uparrow\right>\!=\!\left(\!\frac{\Delta_1\Delta_2}{4\epsilon_1\epsilon_2},-\frac{\Delta_1}{2\epsilon_1},-\frac{\Delta_2}{2\epsilon_2},1-\frac{\Delta_1^2}{8\epsilon_1^2}-\frac{\Delta_2^2}{8\epsilon_2^2}\!\right)^T\!.
\end{aligned}
\label{eq:eigenstates}
\end{equation}
The corresponding eigenenergies are
\begin{equation}
\begin{gathered}
E_1=-\frac{\epsilon_1+\epsilon_2}{2}-\frac{\Delta_1^2}{4\epsilon_1}-\frac{\Delta_2^2}{4\epsilon_2},\\
E_2=-\frac{\epsilon_1-\epsilon_2}{2}-\frac{\Delta_1^2}{4\epsilon_1}+\frac{\Delta_2^2}{4\epsilon_2},\\
E_3=\frac{\epsilon_1-\epsilon_2}{2}+\frac{\Delta_1^2}{4\epsilon_1}-\frac{\Delta_2^2}{4\epsilon_2},\\
E_4=\frac{\epsilon_1+\epsilon_2}{2}+\frac{\Delta_1^2}{4\epsilon_1}+\frac{\Delta_2^2}{4\epsilon_2}.
\end{gathered}
\label{eq:eigenenergies}
\end{equation}
Using Eqs.~\eqref{eq:floquet_solution_unperturbed} and \eqref{eq:floquet_solution_special}, up to second order of tunnel splitting energies $\Delta_{q}$ we obtain following Fourier components, $\left|u_{\alpha k}\right>$, of the Floquet states:
\begin{widetext}
\begin{equation}
\begin{aligned}
\left|u_{1k}\right>\!&=\!
\begin{pmatrix} J_k\left(A/\omega\right)\left(1\!-\!\dfrac{1}{2}\sum\limits_{n=-\infty}^{+\infty}\left(\Delta_1^2\lambda_{1 n}^{+2}\!+\!\Delta_2^2\lambda_{2 n}^{+2}\right)\right)\!+\!\dfrac{1}{2}\sum\limits_{\substack{m=-\infty\\m\ne0}}^{+\infty}J_{k-m}\left(A/\omega\right)\dfrac{\Delta_1^2\chi_{1,-m}^{+}\!+\!\Delta_2^2\chi_{2,-m}^{+}}{m\omega}\\
\Delta_2\lambda_{2 k}^{+}\\
\Delta_1\lambda_{1 k}^{+}\\
\dfrac{\Delta_1\Delta_2}{2}\sum\limits_{n,m=-\infty}^{+\infty}\left(\lambda_{1 n}^{+}\!+\!\lambda_{2 n}^{+}\right)\dfrac{J_{m-k}\left(A/\omega\right)J_{m-n}\left(A/\omega\right)}{\epsilon_1\!+\!\epsilon_2\!+\!m\omega}
\end{pmatrix}
,\\
\gamma_{1}\!&=\!-\frac{\epsilon_{1}+\epsilon_{2}+g}{2}-\frac{1}{2}\left(\Delta_1^2\chi_{10}^{+}+\Delta_2^2\chi_{20}^{+}\right),
\end{aligned}
\label{eq:floquet_u1_gamma_1}
\end{equation}
\begin{equation}
\begin{aligned}
\left|u_{2k}\right>\!&=\!
	\begin{pmatrix}
	-\Delta_2\chi_{2k}^{+}\\
	\delta_{k0}+\begin{cases}
	-\dfrac{1}{2}\sum\limits_{n=-\infty}^{+\infty}\left(\Delta_1^2\lambda_{1 n}^{-2}\!+\!\Delta_2^2\lambda_{2 n}^{+2}\right), & \text{if}\ k=0 \\
	\dfrac{\Delta_1^2\chi_{1k}^{-}\!-\!\Delta_2^2\chi_{2k}^{+}}{2k\omega}, & \text{otherwise}
	\end{cases}\\
	\dfrac{\Delta_1\Delta_2}{2}\dfrac{\chi_{1k}^{-}\!-\!\chi_{2k}^{+}}{\epsilon_1\!-\!\epsilon_2\!+\!k\omega}\\
	\Delta_1\chi_{1k}^{-}
	\end{pmatrix}
,\\
\gamma_{2}\!&=\!-\frac{\epsilon_{1}-\epsilon_{2}-g}{2}-\frac{1}{2}\left(\Delta_1^2\chi_{10}^{-}-\Delta_2^2\chi_{20}^{+}\right),
\end{aligned}
\label{eq:floquet_u2_gamma_2}
\end{equation}
\begin{equation}
\begin{aligned}
\left|u_{3k}\right>\!&=\!
	\begin{pmatrix}
	-\Delta_1\chi_{1k}^{+}\\
	\dfrac{\Delta_1\Delta_2}{2}\dfrac{\chi_{2k}^{-}\!-\!\chi_{1k}^{+}}{\epsilon_2\!-\!\epsilon_1\!+\!k\omega}\\
	\delta_{k0}+\begin{cases}
	-\dfrac{1}{2}\sum\limits_{n=-\infty}^{+\infty}\left(\Delta_1^2\lambda_{1 n}^{+2}\!+\!\Delta_2^2\lambda_{2n}^{-2}\right), & \text{if}\ k=0 \\
	\dfrac{\Delta_2^2\chi_{2k}^{-}\!-\!\Delta_1^2\chi_{1k}^{+}}{2k\omega}, & \text{otherwise}
	\end{cases}\\
	\Delta_2\chi_{2k}^{-}
	\end{pmatrix}
,\\
\gamma_{3}\!&=\!\frac{\epsilon_{1}-\epsilon_{2}+g}{2}+\frac{1}{2}\left(\Delta_1^2\chi_{10}^{+}-\Delta_2^2\chi_{20}^{-}\right),
\end{aligned}
\label{eq:floquet_u3_gamma_3}
\end{equation}
\begin{equation}
\begin{aligned}
\left|u_{4k}\right>\!&=\!
	\begin{pmatrix} 
	-\dfrac{\Delta_1\Delta_2}{2}\sum\limits_{n,m=-\infty}^{+\infty}\left(\lambda_{1 n}^{-}\!+\!\lambda_{2 n}^{-}\right)\dfrac{J_{k+m}\left(A/\omega\right)J_{n+m}\left(A/\omega\right)}{\epsilon_1\!+\!\epsilon_2\!+\!m\omega}\\
	\Delta_1\lambda_{1 k}^{-}\\
	\Delta_2\lambda_{2 k}^{-}\\
	J_{-k}\left(A/\omega\right)\left(1\!-\!\dfrac{1}{2}\sum\limits_{n=-\infty}^{+\infty}\left(\Delta_1^2\lambda_{1 n}^{-2}\!+\!\Delta_2^2\lambda_{2 n}^{-2}\right)\right)\!-\!\dfrac{1}{2}\sum\limits_{\substack{m=-\infty\\m\ne0}}^{+\infty}J_{m-k}\left(A/\omega\right)\dfrac{\Delta_1^2\chi_{1,-m}^{-}\!+\!\Delta_2^2\chi_{2,-m}^{-}}{m\omega}\\
	\end{pmatrix}
,\\
\gamma_{4}\!&=\!\frac{\epsilon_{1}+\epsilon_{2}-g}{2}+\frac{1}{2}\left(\Delta_1^2\chi_{10}^{-}+\Delta_2^2\chi_{20}^{-}\right).
\end{aligned}
\label{eq:floquet_u4_gamma_4}
\end{equation}

According to \eqref{eq:eigenstates} and \eqref{eq:floquet_u1_gamma_1}--\eqref{eq:floquet_u4_gamma_4}, one can obtain the lowest order, in this smallness parameter, matrix elements $S_{\alpha x}$ approximation by keeping only the first term in the infinite series,
\begin{equation}
\begin{gathered}
S_{11}=S_{22}=S_{33}=S_{44}=1,\\
S_{12}=S_{21}=\frac{\Delta_2^2}{4\epsilon_2^2}\sum_{k=-\infty}^{+\infty}\left[J_{k}\left(A/\omega\right)\frac{g\!+\!k\omega}{\epsilon_2\!+\!g\!+\!k\omega}\right]^2,\\
S_{13}=S_{31}=\frac{\Delta_1^2}{4\epsilon_1^2}\sum_{k=-\infty}^{+\infty}\left[J_{k}\left(A/\omega\right)\frac{g\!+\!k\omega}{\epsilon_1\!+\!g\!+\!k\omega}\right]^2,\\
S_{24}=S_{42}=\frac{\Delta_1^2}{4\epsilon_1^2}\sum_{k=-\infty}^{+\infty}\left[J_{k}\left(A/\omega\right)\frac{g\!+\!k\omega}{-\epsilon_1\!+\!g\!+\!k\omega}\right]^2,\\
S_{34}=S_{43}=\frac{\Delta_2^2}{4\epsilon_2^2}\sum_{k=-\infty}^{+\infty}\left[J_{k}\left(A/\omega\right)\frac{g\!+\!k\omega}{-\epsilon_2\!+\!g\!+\!k\omega}\right]^2,\\
S_{14}=\frac{\Delta_1^2\Delta_2^2}{16\epsilon_1^2\epsilon_2^2}\sum_{k=-\infty}^{+\infty}\left[\sum_{n=-\infty}^{+\infty}J_{n}\left(A/\omega\right)J_{k-n}\left(A/\omega\right)\left(1\!-\!\frac{1}{\epsilon_1\!+\!\epsilon_2\!+\!k\omega}\left(\frac{\epsilon_1\!\left(\epsilon_1\!+\!k\omega\right)}{\epsilon_1\!+\!g\!+\!n\omega}\!+\!\frac{\epsilon_2\!\left(\epsilon_2\!+\!k\omega\right)}{\epsilon_2\!+\!g\!+\!n\omega}\right)\right)\right]^2,\\
S_{41}=\frac{\Delta_1^2\Delta_2^2}{16\epsilon_1^2\epsilon_2^2}\sum_{k=-\infty}^{+\infty}\left[\sum_{n=-\infty}^{+\infty}J_{n}\left(A/\omega\right)J_{k-n}\left(A/\omega\right)\left(1\!-\!\frac{1}{\epsilon_1\!+\!\epsilon_2\!+\!k\omega}\left(\frac{\epsilon_1\!\left(\epsilon_1\!+\!k\omega\right)}{\epsilon_1\!-\!g\!+\!n\omega}\!+\!\frac{\epsilon_2\!\left(\epsilon_2\!+\!k\omega\right)}{\epsilon_2\!-\!g\!+\!n\omega}\right)\right)\right]^2.
\end{gathered}
\label{eq:S_matrix_elements}
\end{equation}
Matrix elements $S_{23}$ and $S_{32}$ are fourth order of smallness and out of our interest, so we do not give their values.
\end{widetext}

\bibliographystyle{apsrev4-2}
\bibliography{bibliography}

\begin{thebibliography}{49}%
\makeatletter
\providecommand \@ifxundefined [1]{%
 \@ifx{#1\undefined}
}%
\providecommand \@ifnum [1]{%
 \ifnum #1\expandafter \@firstoftwo
 \else \expandafter \@secondoftwo
 \fi
}%
\providecommand \@ifx [1]{%
 \ifx #1\expandafter \@firstoftwo
 \else \expandafter \@secondoftwo
 \fi
}%
\providecommand \natexlab [1]{#1}%
\providecommand \enquote  [1]{``#1''}%
\providecommand \bibnamefont  [1]{#1}%
\providecommand \bibfnamefont [1]{#1}%
\providecommand \citenamefont [1]{#1}%
\providecommand \href@noop [0]{\@secondoftwo}%
\providecommand \href [0]{\begingroup \@sanitize@url \@href}%
\providecommand \@href[1]{\@@startlink{#1}\@@href}%
\providecommand \@@href[1]{\endgroup#1\@@endlink}%
\providecommand \@sanitize@url [0]{\catcode `\\12\catcode `\$12\catcode
  `\&12\catcode `\#12\catcode `\^12\catcode `\_12\catcode `\%12\relax}%
\providecommand \@@startlink[1]{}%
\providecommand \@@endlink[0]{}%
\providecommand \url  [0]{\begingroup\@sanitize@url \@url }%
\providecommand \@url [1]{\endgroup\@href {#1}{\urlprefix }}%
\providecommand \urlprefix  [0]{URL }%
\providecommand \Eprint [0]{\href }%
\providecommand \doibase [0]{https://doi.org/}%
\providecommand \selectlanguage [0]{\@gobble}%
\providecommand \bibinfo  [0]{\@secondoftwo}%
\providecommand \bibfield  [0]{\@secondoftwo}%
\providecommand \translation [1]{[#1]}%
\providecommand \BibitemOpen [0]{}%
\providecommand \bibitemStop [0]{}%
\providecommand \bibitemNoStop [0]{.\EOS\space}%
\providecommand \EOS [0]{\spacefactor3000\relax}%
\providecommand \BibitemShut  [1]{\csname bibitem#1\endcsname}%
\let\auto@bib@innerbib\@empty
\bibitem [{\citenamefont {Vahala}(2004)}]{Vahala2004}%
  \BibitemOpen
  \bibinfo {editor} {\bibfnamefont {K.}~\bibnamefont {Vahala}},\ ed.,\
  \href@noop {} {\emph {\bibinfo {title} {Optical {Microcavities}}}},\ \bibinfo
  {series} {Advanced Series in Applied Physics}, Vol.~\bibinfo {volume} {5}\
  (\bibinfo  {publisher} {World Scientific},\ \bibinfo {year}
  {2004})\BibitemShut {NoStop}%
\bibitem [{\citenamefont {Wendin}(2017)}]{Wendin2017}%
  \BibitemOpen
  \bibfield  {author} {\bibinfo {author} {\bibfnamefont {G.}~\bibnamefont
  {Wendin}},\ }\href@noop {} {\bibfield  {journal} {\bibinfo  {journal} {Rep.
  Prog. Phys.}\ }\textbf {\bibinfo {volume} {80}},\ \bibinfo {pages} {106001}
  (\bibinfo {year} {2017})}\BibitemShut {NoStop}%
\bibitem [{\citenamefont {Kockum}\ and\ \citenamefont
  {Nori}(2019)}]{Kockum2019}%
  \BibitemOpen
  \bibfield  {author} {\bibinfo {author} {\bibfnamefont {A.~F.}\ \bibnamefont
  {Kockum}}\ and\ \bibinfo {author} {\bibfnamefont {F.}~\bibnamefont {Nori}},\
  }in\ \href@noop {} {\emph {\bibinfo {booktitle} {Fundamentals and Frontiers
  of the Josephson Effect}}},\ \bibinfo {series} {Springer Series in Materials
  Science}, Vol.\ \bibinfo {volume} {286},\ \bibinfo {editor} {edited by\
  \bibinfo {editor} {\bibfnamefont {F.}~\bibnamefont {Tafuri}}}\ (\bibinfo
  {publisher} {Springer},\ \bibinfo {address} {Cham},\ \bibinfo {year} {2019})\
  pp.\ \bibinfo {pages} {703--741}\BibitemShut {NoStop}%
\bibitem [{\citenamefont {Kjaergaard}\ \emph {et~al.}(2020)\citenamefont
  {Kjaergaard}, \citenamefont {Schwartz}, \citenamefont {Braum\"{u}ller},
  \citenamefont {Krantz}, \citenamefont {Wang}, \citenamefont {Gustavsson},\
  and\ \citenamefont {Oliver}}]{Kjaergaard2020}%
  \BibitemOpen
  \bibfield  {author} {\bibinfo {author} {\bibfnamefont {M.}~\bibnamefont
  {Kjaergaard}}, \bibinfo {author} {\bibfnamefont {M.~E.}\ \bibnamefont
  {Schwartz}}, \bibinfo {author} {\bibfnamefont {J.}~\bibnamefont
  {Braum\"{u}ller}}, \bibinfo {author} {\bibfnamefont {P.}~\bibnamefont
  {Krantz}}, \bibinfo {author} {\bibfnamefont {J.~I.-J.}\ \bibnamefont {Wang}},
  \bibinfo {author} {\bibfnamefont {S.}~\bibnamefont {Gustavsson}},\ and\
  \bibinfo {author} {\bibfnamefont {W.~D.}\ \bibnamefont {Oliver}},\
  }\href@noop {} {\bibfield  {journal} {\bibinfo  {journal} {Annu. Rev.
  Condens. Matter Phys.}\ }\textbf {\bibinfo {volume} {11}},\ \bibinfo {pages}
  {369} (\bibinfo {year} {2020})}\BibitemShut {NoStop}%
\bibitem [{\citenamefont {Oelsner}\ \emph {et~al.}(2010)\citenamefont
  {Oelsner}, \citenamefont {van~der Ploeg}, \citenamefont {Macha},
  \citenamefont {H\"{u}bner}, \citenamefont {Born}, \citenamefont {Anders},
  \citenamefont {Il'ichev}, \citenamefont {Meyer}, \citenamefont {Grajcar},
  \citenamefont {W\"{u}nsch}, \citenamefont {Siegel}, \citenamefont
  {Omelyanchouk},\ and\ \citenamefont {Astafiev}}]{Oelsner2010}%
  \BibitemOpen
  \bibfield  {author} {\bibinfo {author} {\bibfnamefont {G.}~\bibnamefont
  {Oelsner}}, \bibinfo {author} {\bibfnamefont {S.~H.~W.}\ \bibnamefont
  {van~der Ploeg}}, \bibinfo {author} {\bibfnamefont {P.}~\bibnamefont
  {Macha}}, \bibinfo {author} {\bibfnamefont {U.}~\bibnamefont {H\"{u}bner}},
  \bibinfo {author} {\bibfnamefont {D.}~\bibnamefont {Born}}, \bibinfo {author}
  {\bibfnamefont {S.}~\bibnamefont {Anders}}, \bibinfo {author} {\bibfnamefont
  {E.}~\bibnamefont {Il'ichev}}, \bibinfo {author} {\bibfnamefont {H.-G.}\
  \bibnamefont {Meyer}}, \bibinfo {author} {\bibfnamefont {M.}~\bibnamefont
  {Grajcar}}, \bibinfo {author} {\bibfnamefont {S.}~\bibnamefont {W\"{u}nsch}},
  \bibinfo {author} {\bibfnamefont {M.}~\bibnamefont {Siegel}}, \bibinfo
  {author} {\bibfnamefont {A.~N.}\ \bibnamefont {Omelyanchouk}},\ and\ \bibinfo
  {author} {\bibfnamefont {O.}~\bibnamefont {Astafiev}},\ }\href@noop {}
  {\bibfield  {journal} {\bibinfo  {journal} {Phys. Rev. B}\ }\textbf {\bibinfo
  {volume} {81}},\ \bibinfo {pages} {172505} (\bibinfo {year}
  {2010})}\BibitemShut {NoStop}%
\bibitem [{\citenamefont {Zagoskin}(2011)}]{Zagoskin2011}%
  \BibitemOpen
  \bibfield  {author} {\bibinfo {author} {\bibfnamefont {A.~M.}\ \bibnamefont
  {Zagoskin}},\ }\href@noop {} {\emph {\bibinfo {title} {Quantum {Engineering}:
  {Theory} and {Design} of {Quantum} {Coherent} {Structures}}}}\ (\bibinfo
  {publisher} {Cambridge University Press},\ \bibinfo {address} {Cambridge},\
  \bibinfo {year} {2011})\BibitemShut {NoStop}%
\bibitem [{\citenamefont {Gu}\ \emph {et~al.}(2017)\citenamefont {Gu},
  \citenamefont {Kockum}, \citenamefont {Miranowicz}, \citenamefont {Liu},\
  and\ \citenamefont {Nori}}]{Gu2017}%
  \BibitemOpen
  \bibfield  {author} {\bibinfo {author} {\bibfnamefont {X.}~\bibnamefont
  {Gu}}, \bibinfo {author} {\bibfnamefont {A.~F.}\ \bibnamefont {Kockum}},
  \bibinfo {author} {\bibfnamefont {A.}~\bibnamefont {Miranowicz}}, \bibinfo
  {author} {\bibfnamefont {Y.}~\bibnamefont {Liu}},\ and\ \bibinfo {author}
  {\bibfnamefont {F.}~\bibnamefont {Nori}},\ }\href@noop {} {\bibfield
  {journal} {\bibinfo  {journal} {Phys. Rep.}\ }\textbf {\bibinfo {volume}
  {718-719}},\ \bibinfo {pages} {1} (\bibinfo {year} {2017})}\BibitemShut
  {NoStop}%
\bibitem [{\citenamefont {Wang}\ \emph {et~al.}(2018)\citenamefont {Wang},
  \citenamefont {Li}, \citenamefont {Yin},\ and\ \citenamefont
  {Zeng}}]{Wang2018}%
  \BibitemOpen
  \bibfield  {author} {\bibinfo {author} {\bibfnamefont {Y.}~\bibnamefont
  {Wang}}, \bibinfo {author} {\bibfnamefont {Y.}~\bibnamefont {Li}}, \bibinfo
  {author} {\bibfnamefont {Z.}~\bibnamefont {Yin}},\ and\ \bibinfo {author}
  {\bibfnamefont {B.}~\bibnamefont {Zeng}},\ }\href@noop {} {\bibfield
  {journal} {\bibinfo  {journal} {npj Quantum Inf.}\ }\textbf {\bibinfo
  {volume} {4}},\ \bibinfo {pages} {46} (\bibinfo {year} {2018})}\BibitemShut
  {NoStop}%
\bibitem [{\citenamefont {Reagor}\ \emph {et~al.}(2018)\citenamefont {Reagor},
  \citenamefont {Osborn}, \citenamefont {Tezak}, \citenamefont {Staley},
  \citenamefont {Prawiroatmodjo}, \citenamefont {Scheer}, \citenamefont
  {Alidoust}, \citenamefont {Sete}, \citenamefont {Didier}, \citenamefont
  {da~Silva}, \citenamefont {Acala}, \citenamefont {Angeles}, \citenamefont
  {Bestwick}, \citenamefont {Block}, \citenamefont {Bloom}, \citenamefont
  {Bradley}, \citenamefont {Bui}, \citenamefont {Caldwell}, \citenamefont
  {Capelluto}, \citenamefont {Chilcott}, \citenamefont {Cordova}, \citenamefont
  {Crossman}, \citenamefont {Curtis}, \citenamefont {Deshpande}, \citenamefont
  {El~Bouayadi}, \citenamefont {Girshovich}, \citenamefont {Hong},
  \citenamefont {Hudson}, \citenamefont {Karalekas}, \citenamefont {Kuang},
  \citenamefont {Lenihan}, \citenamefont {Manenti}, \citenamefont {Manning},
  \citenamefont {Marshall}, \citenamefont {Mohan}, \citenamefont {O'Brien},
  \citenamefont {Otterbach}, \citenamefont {Papageorge}, \citenamefont
  {Paquette}, \citenamefont {Pelstring}, \citenamefont {Polloreno},
  \citenamefont {Rawat}, \citenamefont {Ryan}, \citenamefont {Renzas},
  \citenamefont {Rubin}, \citenamefont {Russel}, \citenamefont {Rust},
  \citenamefont {Scarabelli}, \citenamefont {Selvanayagam}, \citenamefont
  {Sinclair}, \citenamefont {Smith}, \citenamefont {Suska}, \citenamefont {To},
  \citenamefont {Vahidpour}, \citenamefont {Vodrahalli}, \citenamefont
  {Whyland}, \citenamefont {Yadav}, \citenamefont {Zeng},\ and\ \citenamefont
  {Rigetti}}]{Reagor2018}%
  \BibitemOpen
  \bibfield  {author} {\bibinfo {author} {\bibfnamefont {M.}~\bibnamefont
  {Reagor}}, \bibinfo {author} {\bibfnamefont {C.~B.}\ \bibnamefont {Osborn}},
  \bibinfo {author} {\bibfnamefont {N.}~\bibnamefont {Tezak}}, \bibinfo
  {author} {\bibfnamefont {A.}~\bibnamefont {Staley}}, \bibinfo {author}
  {\bibfnamefont {G.}~\bibnamefont {Prawiroatmodjo}}, \bibinfo {author}
  {\bibfnamefont {M.}~\bibnamefont {Scheer}}, \bibinfo {author} {\bibfnamefont
  {N.}~\bibnamefont {Alidoust}}, \bibinfo {author} {\bibfnamefont {E.~A.}\
  \bibnamefont {Sete}}, \bibinfo {author} {\bibfnamefont {N.}~\bibnamefont
  {Didier}}, \bibinfo {author} {\bibfnamefont {M.~P.}\ \bibnamefont
  {da~Silva}}, \bibinfo {author} {\bibfnamefont {E.}~\bibnamefont {Acala}},
  \bibinfo {author} {\bibfnamefont {J.}~\bibnamefont {Angeles}}, \bibinfo
  {author} {\bibfnamefont {A.}~\bibnamefont {Bestwick}}, \bibinfo {author}
  {\bibfnamefont {M.}~\bibnamefont {Block}}, \bibinfo {author} {\bibfnamefont
  {B.}~\bibnamefont {Bloom}}, \bibinfo {author} {\bibfnamefont
  {A.}~\bibnamefont {Bradley}}, \bibinfo {author} {\bibfnamefont
  {C.}~\bibnamefont {Bui}}, \bibinfo {author} {\bibfnamefont {S.}~\bibnamefont
  {Caldwell}}, \bibinfo {author} {\bibfnamefont {L.}~\bibnamefont {Capelluto}},
  \bibinfo {author} {\bibfnamefont {R.}~\bibnamefont {Chilcott}}, \bibinfo
  {author} {\bibfnamefont {J.}~\bibnamefont {Cordova}}, \bibinfo {author}
  {\bibfnamefont {G.}~\bibnamefont {Crossman}}, \bibinfo {author}
  {\bibfnamefont {M.}~\bibnamefont {Curtis}}, \bibinfo {author} {\bibfnamefont
  {S.}~\bibnamefont {Deshpande}}, \bibinfo {author} {\bibfnamefont
  {T.}~\bibnamefont {El~Bouayadi}}, \bibinfo {author} {\bibfnamefont
  {D.}~\bibnamefont {Girshovich}}, \bibinfo {author} {\bibfnamefont
  {S.}~\bibnamefont {Hong}}, \bibinfo {author} {\bibfnamefont {A.}~\bibnamefont
  {Hudson}}, \bibinfo {author} {\bibfnamefont {P.}~\bibnamefont {Karalekas}},
  \bibinfo {author} {\bibfnamefont {K.}~\bibnamefont {Kuang}}, \bibinfo
  {author} {\bibfnamefont {M.}~\bibnamefont {Lenihan}}, \bibinfo {author}
  {\bibfnamefont {R.}~\bibnamefont {Manenti}}, \bibinfo {author} {\bibfnamefont
  {T.}~\bibnamefont {Manning}}, \bibinfo {author} {\bibfnamefont
  {J.}~\bibnamefont {Marshall}}, \bibinfo {author} {\bibfnamefont
  {Y.}~\bibnamefont {Mohan}}, \bibinfo {author} {\bibfnamefont
  {W.}~\bibnamefont {O'Brien}}, \bibinfo {author} {\bibfnamefont
  {J.}~\bibnamefont {Otterbach}}, \bibinfo {author} {\bibfnamefont
  {A.}~\bibnamefont {Papageorge}}, \bibinfo {author} {\bibfnamefont {J.-P.}\
  \bibnamefont {Paquette}}, \bibinfo {author} {\bibfnamefont {M.}~\bibnamefont
  {Pelstring}}, \bibinfo {author} {\bibfnamefont {A.}~\bibnamefont
  {Polloreno}}, \bibinfo {author} {\bibfnamefont {V.}~\bibnamefont {Rawat}},
  \bibinfo {author} {\bibfnamefont {C.~A.}\ \bibnamefont {Ryan}}, \bibinfo
  {author} {\bibfnamefont {R.}~\bibnamefont {Renzas}}, \bibinfo {author}
  {\bibfnamefont {N.}~\bibnamefont {Rubin}}, \bibinfo {author} {\bibfnamefont
  {D.}~\bibnamefont {Russel}}, \bibinfo {author} {\bibfnamefont
  {M.}~\bibnamefont {Rust}}, \bibinfo {author} {\bibfnamefont {D.}~\bibnamefont
  {Scarabelli}}, \bibinfo {author} {\bibfnamefont {M.}~\bibnamefont
  {Selvanayagam}}, \bibinfo {author} {\bibfnamefont {R.}~\bibnamefont
  {Sinclair}}, \bibinfo {author} {\bibfnamefont {R.}~\bibnamefont {Smith}},
  \bibinfo {author} {\bibfnamefont {M.}~\bibnamefont {Suska}}, \bibinfo
  {author} {\bibfnamefont {T.-W.}\ \bibnamefont {To}}, \bibinfo {author}
  {\bibfnamefont {M.}~\bibnamefont {Vahidpour}}, \bibinfo {author}
  {\bibfnamefont {N.}~\bibnamefont {Vodrahalli}}, \bibinfo {author}
  {\bibfnamefont {T.}~\bibnamefont {Whyland}}, \bibinfo {author} {\bibfnamefont
  {K.}~\bibnamefont {Yadav}}, \bibinfo {author} {\bibfnamefont
  {W.}~\bibnamefont {Zeng}},\ and\ \bibinfo {author} {\bibfnamefont {C.~T.}\
  \bibnamefont {Rigetti}},\ }\href@noop {} {\bibfield  {journal} {\bibinfo
  {journal} {Sci. Adv.}\ }\textbf {\bibinfo {volume} {4}},\ \bibinfo {pages}
  {eaao3603} (\bibinfo {year} {2018})}\BibitemShut {NoStop}%
\bibitem [{\citenamefont {Neill}\ \emph {et~al.}(2018)\citenamefont {Neill},
  \citenamefont {Roushan}, \citenamefont {Kechedzhi}, \citenamefont {Boixo},
  \citenamefont {Isakov}, \citenamefont {Smelyanskiy}, \citenamefont {Megrant},
  \citenamefont {Chiaro}, \citenamefont {Dunsworth}, \citenamefont {Arya},
  \citenamefont {Barends}, \citenamefont {Burkett}, \citenamefont {Chen},
  \citenamefont {Chen}, \citenamefont {Fowler}, \citenamefont {Foxen},
  \citenamefont {Giustina}, \citenamefont {Graff}, \citenamefont {Jeffrey},
  \citenamefont {Huang}, \citenamefont {Kelly}, \citenamefont {Klimov},
  \citenamefont {Lucero}, \citenamefont {Mutus}, \citenamefont {Neeley},
  \citenamefont {Quintana}, \citenamefont {Sank}, \citenamefont {Vainsencher},
  \citenamefont {Wenner}, \citenamefont {White}, \citenamefont {Neven},\ and\
  \citenamefont {Martinis}}]{Neill2018}%
  \BibitemOpen
  \bibfield  {author} {\bibinfo {author} {\bibfnamefont {C.}~\bibnamefont
  {Neill}}, \bibinfo {author} {\bibfnamefont {P.}~\bibnamefont {Roushan}},
  \bibinfo {author} {\bibfnamefont {K.}~\bibnamefont {Kechedzhi}}, \bibinfo
  {author} {\bibfnamefont {S.}~\bibnamefont {Boixo}}, \bibinfo {author}
  {\bibfnamefont {S.~V.}\ \bibnamefont {Isakov}}, \bibinfo {author}
  {\bibfnamefont {V.}~\bibnamefont {Smelyanskiy}}, \bibinfo {author}
  {\bibfnamefont {A.}~\bibnamefont {Megrant}}, \bibinfo {author} {\bibfnamefont
  {B.}~\bibnamefont {Chiaro}}, \bibinfo {author} {\bibfnamefont
  {A.}~\bibnamefont {Dunsworth}}, \bibinfo {author} {\bibfnamefont
  {K.}~\bibnamefont {Arya}}, \bibinfo {author} {\bibfnamefont {R.}~\bibnamefont
  {Barends}}, \bibinfo {author} {\bibfnamefont {B.}~\bibnamefont {Burkett}},
  \bibinfo {author} {\bibfnamefont {Y.}~\bibnamefont {Chen}}, \bibinfo {author}
  {\bibfnamefont {Z.}~\bibnamefont {Chen}}, \bibinfo {author} {\bibfnamefont
  {A.}~\bibnamefont {Fowler}}, \bibinfo {author} {\bibfnamefont
  {B.}~\bibnamefont {Foxen}}, \bibinfo {author} {\bibfnamefont
  {M.}~\bibnamefont {Giustina}}, \bibinfo {author} {\bibfnamefont
  {R.}~\bibnamefont {Graff}}, \bibinfo {author} {\bibfnamefont
  {E.}~\bibnamefont {Jeffrey}}, \bibinfo {author} {\bibfnamefont
  {T.}~\bibnamefont {Huang}}, \bibinfo {author} {\bibfnamefont
  {J.}~\bibnamefont {Kelly}}, \bibinfo {author} {\bibfnamefont
  {P.}~\bibnamefont {Klimov}}, \bibinfo {author} {\bibfnamefont
  {E.}~\bibnamefont {Lucero}}, \bibinfo {author} {\bibfnamefont
  {J.}~\bibnamefont {Mutus}}, \bibinfo {author} {\bibfnamefont
  {M.}~\bibnamefont {Neeley}}, \bibinfo {author} {\bibfnamefont
  {C.}~\bibnamefont {Quintana}}, \bibinfo {author} {\bibfnamefont
  {D.}~\bibnamefont {Sank}}, \bibinfo {author} {\bibfnamefont {A.}~\bibnamefont
  {Vainsencher}}, \bibinfo {author} {\bibfnamefont {J.}~\bibnamefont {Wenner}},
  \bibinfo {author} {\bibfnamefont {T.~C.}\ \bibnamefont {White}}, \bibinfo
  {author} {\bibfnamefont {H.}~\bibnamefont {Neven}},\ and\ \bibinfo {author}
  {\bibfnamefont {J.~M.}\ \bibnamefont {Martinis}},\ }\href@noop {} {\bibfield
  {journal} {\bibinfo  {journal} {Science}\ }\textbf {\bibinfo {volume}
  {360}},\ \bibinfo {pages} {195} (\bibinfo {year} {2018})}\BibitemShut
  {NoStop}%
\bibitem [{\citenamefont {Arute}\ \emph {et~al.}(2019)\citenamefont {Arute},
  \citenamefont {Arya}, \citenamefont {Babbush}, \citenamefont {Bacon},
  \citenamefont {Bardin}, \citenamefont {Barends}, \citenamefont {Biswas},
  \citenamefont {Boixo}, \citenamefont {Brandao}, \citenamefont {Buell},
  \citenamefont {Burkett}, \citenamefont {Chen}, \citenamefont {Chen},
  \citenamefont {Chiaro}, \citenamefont {Collins}, \citenamefont {Courtney},
  \citenamefont {Dunsworth}, \citenamefont {Farhi}, \citenamefont {Foxen},
  \citenamefont {Fowler}, \citenamefont {Gidney}, \citenamefont {Giustina},
  \citenamefont {Graff}, \citenamefont {Guerin}, \citenamefont {Habegger},
  \citenamefont {Harrigan}, \citenamefont {Hartmann}, \citenamefont {Ho},
  \citenamefont {Hoffmann}, \citenamefont {Huang}, \citenamefont {Humble},
  \citenamefont {Isakov}, \citenamefont {Jeffrey}, \citenamefont {Jiang},
  \citenamefont {Kafri}, \citenamefont {Kechedzhi}, \citenamefont {Kelly},
  \citenamefont {Klimov}, \citenamefont {Knysh}, \citenamefont {Korotkov},
  \citenamefont {Kostritsa}, \citenamefont {Landhuis}, \citenamefont
  {Lindmark}, \citenamefont {Lucero}, \citenamefont {Lyakh}, \citenamefont
  {Mandr\`{a}}, \citenamefont {McClean}, \citenamefont {McEwen}, \citenamefont
  {Megrant}, \citenamefont {Mi}, \citenamefont {Michielsen}, \citenamefont
  {Mohseni}, \citenamefont {Mutus}, \citenamefont {Naaman}, \citenamefont
  {Neeley}, \citenamefont {Neill}, \citenamefont {Niu}, \citenamefont {Ostby},
  \citenamefont {Petukhov}, \citenamefont {Platt}, \citenamefont {Quintana},
  \citenamefont {Rieffel}, \citenamefont {Roushan}, \citenamefont {Rubin},
  \citenamefont {Sank}, \citenamefont {Satzinger}, \citenamefont {Smelyanskiy},
  \citenamefont {Sung}, \citenamefont {Trevithick}, \citenamefont
  {Vainsencher}, \citenamefont {Villalonga}, \citenamefont {White},
  \citenamefont {Yao}, \citenamefont {Yeh}, \citenamefont {Zalcman},
  \citenamefont {Neven},\ and\ \citenamefont {Martinis}}]{Arute2019}%
  \BibitemOpen
  \bibfield  {author} {\bibinfo {author} {\bibfnamefont {F.}~\bibnamefont
  {Arute}}, \bibinfo {author} {\bibfnamefont {K.}~\bibnamefont {Arya}},
  \bibinfo {author} {\bibfnamefont {R.}~\bibnamefont {Babbush}}, \bibinfo
  {author} {\bibfnamefont {D.}~\bibnamefont {Bacon}}, \bibinfo {author}
  {\bibfnamefont {J.~C.}\ \bibnamefont {Bardin}}, \bibinfo {author}
  {\bibfnamefont {R.}~\bibnamefont {Barends}}, \bibinfo {author} {\bibfnamefont
  {R.}~\bibnamefont {Biswas}}, \bibinfo {author} {\bibfnamefont
  {S.}~\bibnamefont {Boixo}}, \bibinfo {author} {\bibfnamefont {F.~G. S.~L.}\
  \bibnamefont {Brandao}}, \bibinfo {author} {\bibfnamefont {D.~A.}\
  \bibnamefont {Buell}}, \bibinfo {author} {\bibfnamefont {B.}~\bibnamefont
  {Burkett}}, \bibinfo {author} {\bibfnamefont {Y.}~\bibnamefont {Chen}},
  \bibinfo {author} {\bibfnamefont {Z.}~\bibnamefont {Chen}}, \bibinfo {author}
  {\bibfnamefont {B.}~\bibnamefont {Chiaro}}, \bibinfo {author} {\bibfnamefont
  {R.}~\bibnamefont {Collins}}, \bibinfo {author} {\bibfnamefont
  {W.}~\bibnamefont {Courtney}}, \bibinfo {author} {\bibfnamefont
  {A.}~\bibnamefont {Dunsworth}}, \bibinfo {author} {\bibfnamefont
  {E.}~\bibnamefont {Farhi}}, \bibinfo {author} {\bibfnamefont
  {B.}~\bibnamefont {Foxen}}, \bibinfo {author} {\bibfnamefont
  {A.}~\bibnamefont {Fowler}}, \bibinfo {author} {\bibfnamefont
  {C.}~\bibnamefont {Gidney}}, \bibinfo {author} {\bibfnamefont
  {M.}~\bibnamefont {Giustina}}, \bibinfo {author} {\bibfnamefont
  {R.}~\bibnamefont {Graff}}, \bibinfo {author} {\bibfnamefont
  {K.}~\bibnamefont {Guerin}}, \bibinfo {author} {\bibfnamefont
  {S.}~\bibnamefont {Habegger}}, \bibinfo {author} {\bibfnamefont {M.~P.}\
  \bibnamefont {Harrigan}}, \bibinfo {author} {\bibfnamefont {M.~J.}\
  \bibnamefont {Hartmann}}, \bibinfo {author} {\bibfnamefont {A.}~\bibnamefont
  {Ho}}, \bibinfo {author} {\bibfnamefont {M.}~\bibnamefont {Hoffmann}},
  \bibinfo {author} {\bibfnamefont {T.}~\bibnamefont {Huang}}, \bibinfo
  {author} {\bibfnamefont {T.~S.}\ \bibnamefont {Humble}}, \bibinfo {author}
  {\bibfnamefont {S.~V.}\ \bibnamefont {Isakov}}, \bibinfo {author}
  {\bibfnamefont {E.}~\bibnamefont {Jeffrey}}, \bibinfo {author} {\bibfnamefont
  {Z.}~\bibnamefont {Jiang}}, \bibinfo {author} {\bibfnamefont
  {D.}~\bibnamefont {Kafri}}, \bibinfo {author} {\bibfnamefont
  {K.}~\bibnamefont {Kechedzhi}}, \bibinfo {author} {\bibfnamefont
  {J.}~\bibnamefont {Kelly}}, \bibinfo {author} {\bibfnamefont {P.~V.}\
  \bibnamefont {Klimov}}, \bibinfo {author} {\bibfnamefont {S.}~\bibnamefont
  {Knysh}}, \bibinfo {author} {\bibfnamefont {A.}~\bibnamefont {Korotkov}},
  \bibinfo {author} {\bibfnamefont {F.}~\bibnamefont {Kostritsa}}, \bibinfo
  {author} {\bibfnamefont {D.}~\bibnamefont {Landhuis}}, \bibinfo {author}
  {\bibfnamefont {M.}~\bibnamefont {Lindmark}}, \bibinfo {author}
  {\bibfnamefont {E.}~\bibnamefont {Lucero}}, \bibinfo {author} {\bibfnamefont
  {D.}~\bibnamefont {Lyakh}}, \bibinfo {author} {\bibfnamefont
  {S.}~\bibnamefont {Mandr\`{a}}}, \bibinfo {author} {\bibfnamefont {J.~R.}\
  \bibnamefont {McClean}}, \bibinfo {author} {\bibfnamefont {M.}~\bibnamefont
  {McEwen}}, \bibinfo {author} {\bibfnamefont {A.}~\bibnamefont {Megrant}},
  \bibinfo {author} {\bibfnamefont {X.}~\bibnamefont {Mi}}, \bibinfo {author}
  {\bibfnamefont {K.}~\bibnamefont {Michielsen}}, \bibinfo {author}
  {\bibfnamefont {M.}~\bibnamefont {Mohseni}}, \bibinfo {author} {\bibfnamefont
  {J.}~\bibnamefont {Mutus}}, \bibinfo {author} {\bibfnamefont
  {O.}~\bibnamefont {Naaman}}, \bibinfo {author} {\bibfnamefont
  {M.}~\bibnamefont {Neeley}}, \bibinfo {author} {\bibfnamefont
  {C.}~\bibnamefont {Neill}}, \bibinfo {author} {\bibfnamefont {M.~Y.}\
  \bibnamefont {Niu}}, \bibinfo {author} {\bibfnamefont {E.}~\bibnamefont
  {Ostby}}, \bibinfo {author} {\bibfnamefont {A.}~\bibnamefont {Petukhov}},
  \bibinfo {author} {\bibfnamefont {J.~C.}\ \bibnamefont {Platt}}, \bibinfo
  {author} {\bibfnamefont {C.}~\bibnamefont {Quintana}}, \bibinfo {author}
  {\bibfnamefont {E.~G.}\ \bibnamefont {Rieffel}}, \bibinfo {author}
  {\bibfnamefont {P.}~\bibnamefont {Roushan}}, \bibinfo {author} {\bibfnamefont
  {N.~C.}\ \bibnamefont {Rubin}}, \bibinfo {author} {\bibfnamefont
  {D.}~\bibnamefont {Sank}}, \bibinfo {author} {\bibfnamefont {K.~J.}\
  \bibnamefont {Satzinger}}, \bibinfo {author} {\bibfnamefont {V.}~\bibnamefont
  {Smelyanskiy}}, \bibinfo {author} {\bibfnamefont {K.~J.}\ \bibnamefont
  {Sung}}, \bibinfo {author} {\bibfnamefont {M.~D.}\ \bibnamefont
  {Trevithick}}, \bibinfo {author} {\bibfnamefont {A.}~\bibnamefont
  {Vainsencher}}, \bibinfo {author} {\bibfnamefont {B.}~\bibnamefont
  {Villalonga}}, \bibinfo {author} {\bibfnamefont {T.}~\bibnamefont {White}},
  \bibinfo {author} {\bibfnamefont {Z.~J.}\ \bibnamefont {Yao}}, \bibinfo
  {author} {\bibfnamefont {P.}~\bibnamefont {Yeh}}, \bibinfo {author}
  {\bibfnamefont {A.}~\bibnamefont {Zalcman}}, \bibinfo {author} {\bibfnamefont
  {H.}~\bibnamefont {Neven}},\ and\ \bibinfo {author} {\bibfnamefont {J.~M.}\
  \bibnamefont {Martinis}},\ }\href@noop {} {\bibfield  {journal} {\bibinfo
  {journal} {Nature}\ }\textbf {\bibinfo {volume} {574}},\ \bibinfo {pages}
  {505} (\bibinfo {year} {2019})}\BibitemShut {NoStop}%
\bibitem [{\citenamefont {Landau}(1932)}]{Landau1932}%
  \BibitemOpen
  \bibfield  {author} {\bibinfo {author} {\bibfnamefont {L.~D.}\ \bibnamefont
  {Landau}},\ }\href@noop {} {\bibfield  {journal} {\bibinfo  {journal} {Phys.
  Z. Sowjetunion}\ }\textbf {\bibinfo {volume} {2}},\ \bibinfo {pages} {46}
  (\bibinfo {year} {1932})}\BibitemShut {NoStop}%
\bibitem [{\citenamefont {Zener}(1932)}]{Zener1932}%
  \BibitemOpen
  \bibfield  {author} {\bibinfo {author} {\bibfnamefont {C.}~\bibnamefont
  {Zener}},\ }\href@noop {} {\bibfield  {journal} {\bibinfo  {journal} {Proc.
  R. Soc. A}\ }\textbf {\bibinfo {volume} {137}},\ \bibinfo {pages} {696}
  (\bibinfo {year} {1932})}\BibitemShut {NoStop}%
\bibitem [{\citenamefont {St\"{u}ckelberg}(1932)}]{Stuckelberg1932}%
  \BibitemOpen
  \bibfield  {author} {\bibinfo {author} {\bibfnamefont {E.~C.~G.}\
  \bibnamefont {St\"{u}ckelberg}},\ }\href@noop {} {\bibfield  {journal}
  {\bibinfo  {journal} {Helv. Phys. Acta}\ }\textbf {\bibinfo {volume} {5}},\
  \bibinfo {pages} {369} (\bibinfo {year} {1932})}\BibitemShut {NoStop}%
\bibitem [{\citenamefont {Shirley}(1965)}]{Shirley1965}%
  \BibitemOpen
  \bibfield  {author} {\bibinfo {author} {\bibfnamefont {J.~H.}\ \bibnamefont
  {Shirley}},\ }\href@noop {} {\bibfield  {journal} {\bibinfo  {journal} {Phys.
  Rev.}\ }\textbf {\bibinfo {volume} {138}},\ \bibinfo {pages} {B979} (\bibinfo
  {year} {1965})}\BibitemShut {NoStop}%
\bibitem [{\citenamefont {Shevchenko}\ \emph {et~al.}(2010)\citenamefont
  {Shevchenko}, \citenamefont {Ashhab},\ and\ \citenamefont
  {Nori}}]{Shevchenko2010}%
  \BibitemOpen
  \bibfield  {author} {\bibinfo {author} {\bibfnamefont {S.~N.}\ \bibnamefont
  {Shevchenko}}, \bibinfo {author} {\bibfnamefont {S.}~\bibnamefont {Ashhab}},\
  and\ \bibinfo {author} {\bibfnamefont {F.}~\bibnamefont {Nori}},\ }\href@noop
  {} {\bibfield  {journal} {\bibinfo  {journal} {Phys. Rep.}\ }\textbf
  {\bibinfo {volume} {492}},\ \bibinfo {pages} {1} (\bibinfo {year}
  {2010})}\BibitemShut {NoStop}%
\bibitem [{\citenamefont {Oliver}\ \emph {et~al.}(2005)\citenamefont {Oliver},
  \citenamefont {Yu}, \citenamefont {Lee}, \citenamefont {Berggren},
  \citenamefont {Levitov},\ and\ \citenamefont {Orlando}}]{Oliver2005}%
  \BibitemOpen
  \bibfield  {author} {\bibinfo {author} {\bibfnamefont {W.~D.}\ \bibnamefont
  {Oliver}}, \bibinfo {author} {\bibfnamefont {Y.}~\bibnamefont {Yu}}, \bibinfo
  {author} {\bibfnamefont {J.~C.}\ \bibnamefont {Lee}}, \bibinfo {author}
  {\bibfnamefont {K.~K.}\ \bibnamefont {Berggren}}, \bibinfo {author}
  {\bibfnamefont {L.~S.}\ \bibnamefont {Levitov}},\ and\ \bibinfo {author}
  {\bibfnamefont {T.~P.}\ \bibnamefont {Orlando}},\ }\href@noop {} {\bibfield
  {journal} {\bibinfo  {journal} {Science}\ }\textbf {\bibinfo {volume}
  {310}},\ \bibinfo {pages} {1653} (\bibinfo {year} {2005})}\BibitemShut
  {NoStop}%
\bibitem [{\citenamefont {Berns}\ \emph {et~al.}(2006)\citenamefont {Berns},
  \citenamefont {Oliver}, \citenamefont {Valenzuela}, \citenamefont {Shytov},
  \citenamefont {Berggren}, \citenamefont {Levitov},\ and\ \citenamefont
  {Orlando}}]{Berns2006}%
  \BibitemOpen
  \bibfield  {author} {\bibinfo {author} {\bibfnamefont {D.~M.}\ \bibnamefont
  {Berns}}, \bibinfo {author} {\bibfnamefont {W.~D.}\ \bibnamefont {Oliver}},
  \bibinfo {author} {\bibfnamefont {S.~O.}\ \bibnamefont {Valenzuela}},
  \bibinfo {author} {\bibfnamefont {A.~V.}\ \bibnamefont {Shytov}}, \bibinfo
  {author} {\bibfnamefont {K.~K.}\ \bibnamefont {Berggren}}, \bibinfo {author}
  {\bibfnamefont {L.~S.}\ \bibnamefont {Levitov}},\ and\ \bibinfo {author}
  {\bibfnamefont {T.~P.}\ \bibnamefont {Orlando}},\ }\href@noop {} {\bibfield
  {journal} {\bibinfo  {journal} {Phys. Rev. Lett.}\ }\textbf {\bibinfo
  {volume} {97}},\ \bibinfo {pages} {150502} (\bibinfo {year}
  {2006})}\BibitemShut {NoStop}%
\bibitem [{\citenamefont {Rudner}\ \emph {et~al.}(2008)\citenamefont {Rudner},
  \citenamefont {Shytov}, \citenamefont {Levitov}, \citenamefont {Berns},
  \citenamefont {Oliver}, \citenamefont {Valenzuela},\ and\ \citenamefont
  {Orlando}}]{Rudner2008}%
  \BibitemOpen
  \bibfield  {author} {\bibinfo {author} {\bibfnamefont {M.~S.}\ \bibnamefont
  {Rudner}}, \bibinfo {author} {\bibfnamefont {A.~V.}\ \bibnamefont {Shytov}},
  \bibinfo {author} {\bibfnamefont {L.~S.}\ \bibnamefont {Levitov}}, \bibinfo
  {author} {\bibfnamefont {D.~M.}\ \bibnamefont {Berns}}, \bibinfo {author}
  {\bibfnamefont {W.~D.}\ \bibnamefont {Oliver}}, \bibinfo {author}
  {\bibfnamefont {S.~O.}\ \bibnamefont {Valenzuela}},\ and\ \bibinfo {author}
  {\bibfnamefont {T.~P.}\ \bibnamefont {Orlando}},\ }\href@noop {} {\bibfield
  {journal} {\bibinfo  {journal} {Phys. Rev. Lett.}\ }\textbf {\bibinfo
  {volume} {101}},\ \bibinfo {pages} {190502} (\bibinfo {year}
  {2008})}\BibitemShut {NoStop}%
\bibitem [{\citenamefont {Izmalkov}\ \emph {et~al.}(2008)\citenamefont
  {Izmalkov}, \citenamefont {van~der Ploeg}, \citenamefont {Shevchenko},
  \citenamefont {Grajcar}, \citenamefont {Il'ichev}, \citenamefont
  {H\"{u}bner}, \citenamefont {Omelyanchouk},\ and\ \citenamefont
  {Meyer}}]{Izmalkov2008}%
  \BibitemOpen
  \bibfield  {author} {\bibinfo {author} {\bibfnamefont {A.}~\bibnamefont
  {Izmalkov}}, \bibinfo {author} {\bibfnamefont {S.~H.~W.}\ \bibnamefont
  {van~der Ploeg}}, \bibinfo {author} {\bibfnamefont {S.~N.}\ \bibnamefont
  {Shevchenko}}, \bibinfo {author} {\bibfnamefont {M.}~\bibnamefont {Grajcar}},
  \bibinfo {author} {\bibfnamefont {E.}~\bibnamefont {Il'ichev}}, \bibinfo
  {author} {\bibfnamefont {U.}~\bibnamefont {H\"{u}bner}}, \bibinfo {author}
  {\bibfnamefont {A.~N.}\ \bibnamefont {Omelyanchouk}},\ and\ \bibinfo {author}
  {\bibfnamefont {H.-G.}\ \bibnamefont {Meyer}},\ }\href@noop {} {\bibfield
  {journal} {\bibinfo  {journal} {Phys. Rev. Lett.}\ }\textbf {\bibinfo
  {volume} {101}},\ \bibinfo {pages} {017003} (\bibinfo {year}
  {2008})}\BibitemShut {NoStop}%
\bibitem [{\citenamefont {Neilinger}\ \emph {et~al.}(2016)\citenamefont
  {Neilinger}, \citenamefont {Shevchenko}, \citenamefont {Bog\'{a}r},
  \citenamefont {Reh\'{a}k}, \citenamefont {Oelsner}, \citenamefont {Karpov},
  \citenamefont {H\"{u}bner}, \citenamefont {Astafiev}, \citenamefont
  {Grajcar},\ and\ \citenamefont {Il’ichev}}]{Neilinger2016}%
  \BibitemOpen
  \bibfield  {author} {\bibinfo {author} {\bibfnamefont {P.}~\bibnamefont
  {Neilinger}}, \bibinfo {author} {\bibfnamefont {S.~N.}\ \bibnamefont
  {Shevchenko}}, \bibinfo {author} {\bibfnamefont {J.}~\bibnamefont
  {Bog\'{a}r}}, \bibinfo {author} {\bibfnamefont {M.}~\bibnamefont
  {Reh\'{a}k}}, \bibinfo {author} {\bibfnamefont {G.}~\bibnamefont {Oelsner}},
  \bibinfo {author} {\bibfnamefont {D.~S.}\ \bibnamefont {Karpov}}, \bibinfo
  {author} {\bibfnamefont {U.}~\bibnamefont {H\"{u}bner}}, \bibinfo {author}
  {\bibfnamefont {O.}~\bibnamefont {Astafiev}}, \bibinfo {author}
  {\bibfnamefont {M.}~\bibnamefont {Grajcar}},\ and\ \bibinfo {author}
  {\bibfnamefont {E.}~\bibnamefont {Il’ichev}},\ }\href@noop {} {\bibfield
  {journal} {\bibinfo  {journal} {Phys. Rev. B}\ }\textbf {\bibinfo {volume}
  {94}},\ \bibinfo {pages} {094519} (\bibinfo {year} {2016})}\BibitemShut
  {NoStop}%
\bibitem [{\citenamefont {Sillanp\"{a}\"{a}}\ \emph {et~al.}(2006)\citenamefont
  {Sillanp\"{a}\"{a}}, \citenamefont {Lehtinen}, \citenamefont {Paila},
  \citenamefont {Makhlin},\ and\ \citenamefont {Hakonen}}]{Sillanpaa2006}%
  \BibitemOpen
  \bibfield  {author} {\bibinfo {author} {\bibfnamefont {M.}~\bibnamefont
  {Sillanp\"{a}\"{a}}}, \bibinfo {author} {\bibfnamefont {T.}~\bibnamefont
  {Lehtinen}}, \bibinfo {author} {\bibfnamefont {A.}~\bibnamefont {Paila}},
  \bibinfo {author} {\bibfnamefont {Y.}~\bibnamefont {Makhlin}},\ and\ \bibinfo
  {author} {\bibfnamefont {P.}~\bibnamefont {Hakonen}},\ }\href@noop {}
  {\bibfield  {journal} {\bibinfo  {journal} {Phys. Rev. Lett.}\ }\textbf
  {\bibinfo {volume} {96}},\ \bibinfo {pages} {187002} (\bibinfo {year}
  {2006})}\BibitemShut {NoStop}%
\bibitem [{\citenamefont {Ribeiro}\ \emph {et~al.}(2013)\citenamefont
  {Ribeiro}, \citenamefont {Petta},\ and\ \citenamefont
  {Burkard}}]{Ribeiro2013}%
  \BibitemOpen
  \bibfield  {author} {\bibinfo {author} {\bibfnamefont {H.}~\bibnamefont
  {Ribeiro}}, \bibinfo {author} {\bibfnamefont {J.~R.}\ \bibnamefont {Petta}},\
  and\ \bibinfo {author} {\bibfnamefont {G.}~\bibnamefont {Burkard}},\
  }\href@noop {} {\bibfield  {journal} {\bibinfo  {journal} {Phys. Rev. B}\
  }\textbf {\bibinfo {volume} {87}},\ \bibinfo {pages} {235318} (\bibinfo
  {year} {2013})}\BibitemShut {NoStop}%
\bibitem [{\citenamefont {Mi}\ \emph {et~al.}(2018)\citenamefont {Mi},
  \citenamefont {Kohler},\ and\ \citenamefont {Petta}}]{Kohler2018}%
  \BibitemOpen
  \bibfield  {author} {\bibinfo {author} {\bibfnamefont {X.}~\bibnamefont
  {Mi}}, \bibinfo {author} {\bibfnamefont {S.}~\bibnamefont {Kohler}},\ and\
  \bibinfo {author} {\bibfnamefont {J.~R.}\ \bibnamefont {Petta}},\ }\href@noop
  {} {\bibfield  {journal} {\bibinfo  {journal} {Phys. Rev. B}\ }\textbf
  {\bibinfo {volume} {98}},\ \bibinfo {pages} {161404(R)} (\bibinfo {year}
  {2018})}\BibitemShut {NoStop}%
\bibitem [{\citenamefont {Blattmann}\ \emph {et~al.}(2015)\citenamefont
  {Blattmann}, \citenamefont {H\"{a}nggi},\ and\ \citenamefont
  {Kohler}}]{Blattmann2015}%
  \BibitemOpen
  \bibfield  {author} {\bibinfo {author} {\bibfnamefont {R.}~\bibnamefont
  {Blattmann}}, \bibinfo {author} {\bibfnamefont {P.}~\bibnamefont
  {H\"{a}nggi}},\ and\ \bibinfo {author} {\bibfnamefont {S.}~\bibnamefont
  {Kohler}},\ }\href@noop {} {\bibfield  {journal} {\bibinfo  {journal} {Phys.
  Rev. A}\ }\textbf {\bibinfo {volume} {91}},\ \bibinfo {pages} {042109}
  (\bibinfo {year} {2015})}\BibitemShut {NoStop}%
\bibitem [{\citenamefont {Quintana}\ \emph {et~al.}(2013)\citenamefont
  {Quintana}, \citenamefont {Petersson}, \citenamefont {McFaul}, \citenamefont
  {Srinivasan}, \citenamefont {Houck},\ and\ \citenamefont
  {Petta}}]{Quintana2013}%
  \BibitemOpen
  \bibfield  {author} {\bibinfo {author} {\bibfnamefont {C.~M.}\ \bibnamefont
  {Quintana}}, \bibinfo {author} {\bibfnamefont {K.~D.}\ \bibnamefont
  {Petersson}}, \bibinfo {author} {\bibfnamefont {L.~W.}\ \bibnamefont
  {McFaul}}, \bibinfo {author} {\bibfnamefont {S.~J.}\ \bibnamefont
  {Srinivasan}}, \bibinfo {author} {\bibfnamefont {A.~A.}\ \bibnamefont
  {Houck}},\ and\ \bibinfo {author} {\bibfnamefont {J.~R.}\ \bibnamefont
  {Petta}},\ }\href@noop {} {\bibfield  {journal} {\bibinfo  {journal} {Phys.
  Rev. Lett.}\ }\textbf {\bibinfo {volume} {110}},\ \bibinfo {pages} {173603}
  (\bibinfo {year} {2013})}\BibitemShut {NoStop}%
\bibitem [{\citenamefont {Roch}\ \emph {et~al.}(2014)\citenamefont {Roch},
  \citenamefont {Schwartz}, \citenamefont {Motzoi}, \citenamefont {Macklin},
  \citenamefont {Vijay}, \citenamefont {Eddins}, \citenamefont {Korotkov},
  \citenamefont {Whaley}, \citenamefont {Sarovar},\ and\ \citenamefont
  {Siddiqi}}]{Roch2014}%
  \BibitemOpen
  \bibfield  {author} {\bibinfo {author} {\bibfnamefont {N.}~\bibnamefont
  {Roch}}, \bibinfo {author} {\bibfnamefont {M.~E.}\ \bibnamefont {Schwartz}},
  \bibinfo {author} {\bibfnamefont {F.}~\bibnamefont {Motzoi}}, \bibinfo
  {author} {\bibfnamefont {C.}~\bibnamefont {Macklin}}, \bibinfo {author}
  {\bibfnamefont {R.}~\bibnamefont {Vijay}}, \bibinfo {author} {\bibfnamefont
  {A.~W.}\ \bibnamefont {Eddins}}, \bibinfo {author} {\bibfnamefont {A.~N.}\
  \bibnamefont {Korotkov}}, \bibinfo {author} {\bibfnamefont {K.~B.}\
  \bibnamefont {Whaley}}, \bibinfo {author} {\bibfnamefont {M.}~\bibnamefont
  {Sarovar}},\ and\ \bibinfo {author} {\bibfnamefont {I.}~\bibnamefont
  {Siddiqi}},\ }\href@noop {} {\bibfield  {journal} {\bibinfo  {journal} {Phys.
  Rev. Lett.}\ }\textbf {\bibinfo {volume} {112}},\ \bibinfo {pages} {170501}
  (\bibinfo {year} {2014})}\BibitemShut {NoStop}%
\bibitem [{\citenamefont {Il'ichev}\ \emph {et~al.}(2010)\citenamefont
  {Il'ichev}, \citenamefont {Shevchenko}, \citenamefont {van~der Ploeg},
  \citenamefont {Grajcar}, \citenamefont {Temchenko}, \citenamefont
  {Omelyanchouk},\ and\ \citenamefont {Meyer}}]{Ilichev2010}%
  \BibitemOpen
  \bibfield  {author} {\bibinfo {author} {\bibfnamefont {E.}~\bibnamefont
  {Il'ichev}}, \bibinfo {author} {\bibfnamefont {S.~N.}\ \bibnamefont
  {Shevchenko}}, \bibinfo {author} {\bibfnamefont {S.~H.~W.}\ \bibnamefont
  {van~der Ploeg}}, \bibinfo {author} {\bibfnamefont {M.}~\bibnamefont
  {Grajcar}}, \bibinfo {author} {\bibfnamefont {E.~A.}\ \bibnamefont
  {Temchenko}}, \bibinfo {author} {\bibfnamefont {A.~N.}\ \bibnamefont
  {Omelyanchouk}},\ and\ \bibinfo {author} {\bibfnamefont {H.-G.}\ \bibnamefont
  {Meyer}},\ }\href@noop {} {\bibfield  {journal} {\bibinfo  {journal} {Phys.
  Rev. B}\ }\textbf {\bibinfo {volume} {81}},\ \bibinfo {pages} {012506}
  (\bibinfo {year} {2010})}\BibitemShut {NoStop}%
\bibitem [{\citenamefont {Majer}\ \emph {et~al.}(2005)\citenamefont {Majer},
  \citenamefont {Paauw}, \citenamefont {ter Haar}, \citenamefont {Harmans},\
  and\ \citenamefont {Mooij}}]{Majer2005}%
  \BibitemOpen
  \bibfield  {author} {\bibinfo {author} {\bibfnamefont {J.~B.}\ \bibnamefont
  {Majer}}, \bibinfo {author} {\bibfnamefont {F.~G.}\ \bibnamefont {Paauw}},
  \bibinfo {author} {\bibfnamefont {A.~C.~J.}\ \bibnamefont {ter Haar}},
  \bibinfo {author} {\bibfnamefont {C.~J. P.~M.}\ \bibnamefont {Harmans}},\
  and\ \bibinfo {author} {\bibfnamefont {J.~E.}\ \bibnamefont {Mooij}},\
  }\href@noop {} {\bibfield  {journal} {\bibinfo  {journal} {Phys. Rev. Lett.}\
  }\textbf {\bibinfo {volume} {94}},\ \bibinfo {pages} {090501} (\bibinfo
  {year} {2005})}\BibitemShut {NoStop}%
\bibitem [{\citenamefont {Weber}\ \emph {et~al.}(2017)\citenamefont {Weber},
  \citenamefont {Samach}, \citenamefont {Hover}, \citenamefont {Gustavsson},
  \citenamefont {Kim}, \citenamefont {Melville}, \citenamefont {Rosenberg},
  \citenamefont {Sears}, \citenamefont {Yan}, \citenamefont {Yoder},
  \citenamefont {Oliver},\ and\ \citenamefont {Kerman}}]{Weber2017}%
  \BibitemOpen
  \bibfield  {author} {\bibinfo {author} {\bibfnamefont {S.~J.}\ \bibnamefont
  {Weber}}, \bibinfo {author} {\bibfnamefont {G.~O.}\ \bibnamefont {Samach}},
  \bibinfo {author} {\bibfnamefont {D.}~\bibnamefont {Hover}}, \bibinfo
  {author} {\bibfnamefont {S.}~\bibnamefont {Gustavsson}}, \bibinfo {author}
  {\bibfnamefont {D.~K.}\ \bibnamefont {Kim}}, \bibinfo {author} {\bibfnamefont
  {A.}~\bibnamefont {Melville}}, \bibinfo {author} {\bibfnamefont
  {D.}~\bibnamefont {Rosenberg}}, \bibinfo {author} {\bibfnamefont {A.~P.}\
  \bibnamefont {Sears}}, \bibinfo {author} {\bibfnamefont {F.}~\bibnamefont
  {Yan}}, \bibinfo {author} {\bibfnamefont {J.~L.}\ \bibnamefont {Yoder}},
  \bibinfo {author} {\bibfnamefont {W.~D.}\ \bibnamefont {Oliver}},\ and\
  \bibinfo {author} {\bibfnamefont {A.~J.}\ \bibnamefont {Kerman}},\
  }\href@noop {} {\bibfield  {journal} {\bibinfo  {journal} {Phys. Rev. Appl.}\
  }\textbf {\bibinfo {volume} {8}},\ \bibinfo {pages} {014004} (\bibinfo {year}
  {2017})}\BibitemShut {NoStop}%
\bibitem [{\citenamefont {Grifoni}\ and\ \citenamefont
  {H\"{a}nggi}(1998)}]{Grifoni1998}%
  \BibitemOpen
  \bibfield  {author} {\bibinfo {author} {\bibfnamefont {M.}~\bibnamefont
  {Grifoni}}\ and\ \bibinfo {author} {\bibfnamefont {P.}~\bibnamefont
  {H\"{a}nggi}},\ }\href@noop {} {\bibfield  {journal} {\bibinfo  {journal}
  {Phys. Rep.}\ }\textbf {\bibinfo {volume} {304}},\ \bibinfo {pages} {229}
  (\bibinfo {year} {1998})}\BibitemShut {NoStop}%
\bibitem [{\citenamefont {Zel'dovich}(1967)}]{Zeldovich1967}%
  \BibitemOpen
  \bibfield  {author} {\bibinfo {author} {\bibfnamefont {Y.~B.}\ \bibnamefont
  {Zel'dovich}},\ }\href@noop {} {\bibfield  {journal} {\bibinfo  {journal}
  {Sov. Phys. JETP}\ }\textbf {\bibinfo {volume} {24}},\ \bibinfo {pages}
  {1006} (\bibinfo {year} {1967})}\BibitemShut {NoStop}%
\bibitem [{\citenamefont {Ritus}(1967)}]{Ritus1967}%
  \BibitemOpen
  \bibfield  {author} {\bibinfo {author} {\bibfnamefont {V.~I.}\ \bibnamefont
  {Ritus}},\ }\href@noop {} {\bibfield  {journal} {\bibinfo  {journal} {Sov.
  Phys. JETP}\ }\textbf {\bibinfo {volume} {24}},\ \bibinfo {pages} {1041}
  (\bibinfo {year} {1967})}\BibitemShut {NoStop}%
\bibitem [{\citenamefont {Sambe}(1973)}]{Sambe1973}%
  \BibitemOpen
  \bibfield  {author} {\bibinfo {author} {\bibfnamefont {H.}~\bibnamefont
  {Sambe}},\ }\href@noop {} {\bibfield  {journal} {\bibinfo  {journal} {Phys.
  Rev. A}\ }\textbf {\bibinfo {volume} {7}},\ \bibinfo {pages} {2203} (\bibinfo
  {year} {1973})}\BibitemShut {NoStop}%
\bibitem [{\citenamefont {Hazewinkel}(1991)}]{Hazewinkel1991}%
  \BibitemOpen
  \bibinfo {editor} {\bibfnamefont {M.}~\bibnamefont {Hazewinkel}},\ ed.,\
  \href@noop {} {\emph {\bibinfo {title} {Encyclopaedia of {Mathematics}:
  Orb-Ray}}}\ (\bibinfo  {publisher} {Kluwer},\ \bibinfo {year}
  {1991})\BibitemShut {NoStop}%
\bibitem [{\citenamefont {Denisenko}\ \emph {et~al.}(2010)\citenamefont
  {Denisenko}, \citenamefont {Satanin}, \citenamefont {Ashhab},\ and\
  \citenamefont {Nori}}]{Satanin2010}%
  \BibitemOpen
  \bibfield  {author} {\bibinfo {author} {\bibfnamefont {M.~V.}\ \bibnamefont
  {Denisenko}}, \bibinfo {author} {\bibfnamefont {A.~M.}\ \bibnamefont
  {Satanin}}, \bibinfo {author} {\bibfnamefont {S.}~\bibnamefont {Ashhab}},\
  and\ \bibinfo {author} {\bibfnamefont {F.}~\bibnamefont {Nori}},\ }\href@noop
  {} {\bibfield  {journal} {\bibinfo  {journal} {Phys. Solid State}\ }\textbf
  {\bibinfo {volume} {52}},\ \bibinfo {pages} {2281} (\bibinfo {year}
  {2010})}\BibitemShut {NoStop}%
\bibitem [{\citenamefont {Denisenko}\ \emph {et~al.}(2012)\citenamefont
  {Denisenko}, \citenamefont {Satanin}, \citenamefont {Ashhab},\ and\
  \citenamefont {Nori}}]{Satanin2012}%
  \BibitemOpen
  \bibfield  {author} {\bibinfo {author} {\bibfnamefont {M.~V.}\ \bibnamefont
  {Denisenko}}, \bibinfo {author} {\bibfnamefont {A.~M.}\ \bibnamefont
  {Satanin}}, \bibinfo {author} {\bibfnamefont {S.}~\bibnamefont {Ashhab}},\
  and\ \bibinfo {author} {\bibfnamefont {F.}~\bibnamefont {Nori}},\ }\href@noop
  {} {\bibfield  {journal} {\bibinfo  {journal} {Phys. Rev. B}\ }\textbf
  {\bibinfo {volume} {85}},\ \bibinfo {pages} {184524} (\bibinfo {year}
  {2012})}\BibitemShut {NoStop}%
\bibitem [{\citenamefont {Sauer}\ \emph {et~al.}(2012)\citenamefont {Sauer},
  \citenamefont {Mintert}, \citenamefont {Gneiting},\ and\ \citenamefont
  {Buchleitner}}]{Sauer2012}%
  \BibitemOpen
  \bibfield  {author} {\bibinfo {author} {\bibfnamefont {S.}~\bibnamefont
  {Sauer}}, \bibinfo {author} {\bibfnamefont {F.}~\bibnamefont {Mintert}},
  \bibinfo {author} {\bibfnamefont {C.}~\bibnamefont {Gneiting}},\ and\
  \bibinfo {author} {\bibfnamefont {A.}~\bibnamefont {Buchleitner}},\
  }\href@noop {} {\bibfield  {journal} {\bibinfo  {journal} {J. Phys. B: At.
  Mol. Opt. Phys.}\ }\textbf {\bibinfo {volume} {45}},\ \bibinfo {pages}
  {154011} (\bibinfo {year} {2012})}\BibitemShut {NoStop}%
\bibitem [{\citenamefont {Gramajo}\ \emph {et~al.}(2017)\citenamefont
  {Gramajo}, \citenamefont {Dom\'{i}nguez},\ and\ \citenamefont
  {S\'{a}nchez}}]{Gramajo2017}%
  \BibitemOpen
  \bibfield  {author} {\bibinfo {author} {\bibfnamefont {A.~L.}\ \bibnamefont
  {Gramajo}}, \bibinfo {author} {\bibfnamefont {D.}~\bibnamefont
  {Dom\'{i}nguez}},\ and\ \bibinfo {author} {\bibfnamefont {M.~J.}\
  \bibnamefont {S\'{a}nchez}},\ }\href@noop {} {\bibfield  {journal} {\bibinfo
  {journal} {Eur. Phys. J. B}\ }\textbf {\bibinfo {volume} {90}},\ \bibinfo
  {pages} {255} (\bibinfo {year} {2017})}\BibitemShut {NoStop}%
\bibitem [{\citenamefont {Gramajo}\ \emph {et~al.}(2018)\citenamefont
  {Gramajo}, \citenamefont {Dom\'{i}nguez},\ and\ \citenamefont
  {S\'{a}nchez}}]{Gramajo2018}%
  \BibitemOpen
  \bibfield  {author} {\bibinfo {author} {\bibfnamefont {A.~L.}\ \bibnamefont
  {Gramajo}}, \bibinfo {author} {\bibfnamefont {D.}~\bibnamefont
  {Dom\'{i}nguez}},\ and\ \bibinfo {author} {\bibfnamefont {M.~J.}\
  \bibnamefont {S\'{a}nchez}},\ }\href@noop {} {\bibfield  {journal} {\bibinfo
  {journal} {Phys. Rev. A}\ }\textbf {\bibinfo {volume} {98}},\ \bibinfo
  {pages} {042337} (\bibinfo {year} {2018})}\BibitemShut {NoStop}%
\bibitem [{\citenamefont {Plourde}\ \emph {et~al.}(2004)\citenamefont
  {Plourde}, \citenamefont {Zhang}, \citenamefont {Whaley}, \citenamefont
  {Wilhelm}, \citenamefont {Robertson}, \citenamefont {Hime}, \citenamefont
  {Linzen}, \citenamefont {Reichardt}, \citenamefont {Wu},\ and\ \citenamefont
  {Clarke}}]{Plourde2004}%
  \BibitemOpen
  \bibfield  {author} {\bibinfo {author} {\bibfnamefont {B.~L.~T.}\
  \bibnamefont {Plourde}}, \bibinfo {author} {\bibfnamefont {J.}~\bibnamefont
  {Zhang}}, \bibinfo {author} {\bibfnamefont {K.~B.}\ \bibnamefont {Whaley}},
  \bibinfo {author} {\bibfnamefont {F.~K.}\ \bibnamefont {Wilhelm}}, \bibinfo
  {author} {\bibfnamefont {T.~L.}\ \bibnamefont {Robertson}}, \bibinfo {author}
  {\bibfnamefont {T.}~\bibnamefont {Hime}}, \bibinfo {author} {\bibfnamefont
  {S.}~\bibnamefont {Linzen}}, \bibinfo {author} {\bibfnamefont {P.~A.}\
  \bibnamefont {Reichardt}}, \bibinfo {author} {\bibfnamefont {C.-E.}\
  \bibnamefont {Wu}},\ and\ \bibinfo {author} {\bibfnamefont {J.}~\bibnamefont
  {Clarke}},\ }\href@noop {} {\bibfield  {journal} {\bibinfo  {journal} {Phys.
  Rev. B}\ }\textbf {\bibinfo {volume} {70}},\ \bibinfo {pages} {140501(R)}
  (\bibinfo {year} {2004})}\BibitemShut {NoStop}%
\bibitem [{\citenamefont {van~der Ploeg}\ \emph {et~al.}(2007)\citenamefont
  {van~der Ploeg}, \citenamefont {Izmalkov}, \citenamefont {van~den Brink},
  \citenamefont {H\"{u}bner}, \citenamefont {Grajcar}, \citenamefont
  {Il'ichev}, \citenamefont {Meyer},\ and\ \citenamefont
  {Zagoskin}}]{Ploeg2007}%
  \BibitemOpen
  \bibfield  {author} {\bibinfo {author} {\bibfnamefont {S.~H.~W.}\
  \bibnamefont {van~der Ploeg}}, \bibinfo {author} {\bibfnamefont
  {A.}~\bibnamefont {Izmalkov}}, \bibinfo {author} {\bibfnamefont {A.~M.}\
  \bibnamefont {van~den Brink}}, \bibinfo {author} {\bibfnamefont
  {U.}~\bibnamefont {H\"{u}bner}}, \bibinfo {author} {\bibfnamefont
  {M.}~\bibnamefont {Grajcar}}, \bibinfo {author} {\bibfnamefont
  {E.}~\bibnamefont {Il'ichev}}, \bibinfo {author} {\bibfnamefont {H.-G.}\
  \bibnamefont {Meyer}},\ and\ \bibinfo {author} {\bibfnamefont {A.~M.}\
  \bibnamefont {Zagoskin}},\ }\href@noop {} {\bibfield  {journal} {\bibinfo
  {journal} {Phys. Rev. Lett.}\ }\textbf {\bibinfo {volume} {98}},\ \bibinfo
  {pages} {057004} (\bibinfo {year} {2007})}\BibitemShut {NoStop}%
\bibitem [{\citenamefont {Groszkowski}\ \emph {et~al.}(2011)\citenamefont
  {Groszkowski}, \citenamefont {Fowler}, \citenamefont {Motzoi},\ and\
  \citenamefont {Wilhelm}}]{Groszkowski2011}%
  \BibitemOpen
  \bibfield  {author} {\bibinfo {author} {\bibfnamefont {P.}~\bibnamefont
  {Groszkowski}}, \bibinfo {author} {\bibfnamefont {A.~G.}\ \bibnamefont
  {Fowler}}, \bibinfo {author} {\bibfnamefont {F.}~\bibnamefont {Motzoi}},\
  and\ \bibinfo {author} {\bibfnamefont {F.~K.}\ \bibnamefont {Wilhelm}},\
  }\href@noop {} {\bibfield  {journal} {\bibinfo  {journal} {Phys. Rev. B}\
  }\textbf {\bibinfo {volume} {84}},\ \bibinfo {pages} {144516} (\bibinfo
  {year} {2011})}\BibitemShut {NoStop}%
\bibitem [{\citenamefont {Allman}\ \emph {et~al.}(2014)\citenamefont {Allman},
  \citenamefont {Whittaker}, \citenamefont {Castellanos-Beltran}, \citenamefont
  {Cicak}, \citenamefont {da~Silva}, \citenamefont {DeFeo}, \citenamefont
  {Lecocq}, \citenamefont {Sirois}, \citenamefont {Teufel}, \citenamefont
  {Aumentado},\ and\ \citenamefont {Simmonds}}]{Allman2014}%
  \BibitemOpen
  \bibfield  {author} {\bibinfo {author} {\bibfnamefont {M.~S.}\ \bibnamefont
  {Allman}}, \bibinfo {author} {\bibfnamefont {J.~D.}\ \bibnamefont
  {Whittaker}}, \bibinfo {author} {\bibfnamefont {M.}~\bibnamefont
  {Castellanos-Beltran}}, \bibinfo {author} {\bibfnamefont {K.}~\bibnamefont
  {Cicak}}, \bibinfo {author} {\bibfnamefont {F.}~\bibnamefont {da~Silva}},
  \bibinfo {author} {\bibfnamefont {M.~P.}\ \bibnamefont {DeFeo}}, \bibinfo
  {author} {\bibfnamefont {F.}~\bibnamefont {Lecocq}}, \bibinfo {author}
  {\bibfnamefont {A.}~\bibnamefont {Sirois}}, \bibinfo {author} {\bibfnamefont
  {J.~D.}\ \bibnamefont {Teufel}}, \bibinfo {author} {\bibfnamefont
  {J.}~\bibnamefont {Aumentado}},\ and\ \bibinfo {author} {\bibfnamefont
  {R.~W.}\ \bibnamefont {Simmonds}},\ }\href@noop {} {\bibfield  {journal}
  {\bibinfo  {journal} {Phys. Rev. Lett.}\ }\textbf {\bibinfo {volume} {112}},\
  \bibinfo {pages} {123601} (\bibinfo {year} {2014})}\BibitemShut {NoStop}%
\bibitem [{\citenamefont {Kohler}\ \emph {et~al.}(1997)\citenamefont {Kohler},
  \citenamefont {Dittrich},\ and\ \citenamefont {H\"{a}nggi}}]{Kohler1997}%
  \BibitemOpen
  \bibfield  {author} {\bibinfo {author} {\bibfnamefont {S.}~\bibnamefont
  {Kohler}}, \bibinfo {author} {\bibfnamefont {T.}~\bibnamefont {Dittrich}},\
  and\ \bibinfo {author} {\bibfnamefont {P.}~\bibnamefont {H\"{a}nggi}},\
  }\href@noop {} {\bibfield  {journal} {\bibinfo  {journal} {Phys. Rev. E}\
  }\textbf {\bibinfo {volume} {55}},\ \bibinfo {pages} {300} (\bibinfo {year}
  {1997})}\BibitemShut {NoStop}%
\bibitem [{\citenamefont {Hone}\ \emph {et~al.}(2009)\citenamefont {Hone},
  \citenamefont {Ketzmerick},\ and\ \citenamefont {Kohn}}]{Hone2009}%
  \BibitemOpen
  \bibfield  {author} {\bibinfo {author} {\bibfnamefont {D.~W.}\ \bibnamefont
  {Hone}}, \bibinfo {author} {\bibfnamefont {R.}~\bibnamefont {Ketzmerick}},\
  and\ \bibinfo {author} {\bibfnamefont {W.}~\bibnamefont {Kohn}},\ }\href@noop
  {} {\bibfield  {journal} {\bibinfo  {journal} {Phys. Rev. E}\ }\textbf
  {\bibinfo {volume} {79}},\ \bibinfo {pages} {051129} (\bibinfo {year}
  {2009})}\BibitemShut {NoStop}%
\bibitem [{\citenamefont {Wootters}(2001)}]{Wootters1998}%
  \BibitemOpen
  \bibfield  {author} {\bibinfo {author} {\bibfnamefont {W.~K.}\ \bibnamefont
  {Wootters}},\ }\href@noop {} {\bibfield  {journal} {\bibinfo  {journal}
  {Quantum Inf. Comput.}\ }\textbf {\bibinfo {volume} {1}},\ \bibinfo {pages}
  {27} (\bibinfo {year} {2001})}\BibitemShut {NoStop}%
\bibitem [{\citenamefont {Gel'man}\ and\ \citenamefont
  {Satanin}(2010)}]{Gelman2010}%
  \BibitemOpen
  \bibfield  {author} {\bibinfo {author} {\bibfnamefont {A.~I.}\ \bibnamefont
  {Gel'man}}\ and\ \bibinfo {author} {\bibfnamefont {A.~M.}\ \bibnamefont
  {Satanin}},\ }\href@noop {} {\bibfield  {journal} {\bibinfo  {journal} {JETP
  Lett.}\ }\textbf {\bibinfo {volume} {91}},\ \bibinfo {pages} {535} (\bibinfo
  {year} {2010})}\BibitemShut {NoStop}%
\bibitem [{\citenamefont {Satanin}\ \emph {et~al.}(2014)\citenamefont
  {Satanin}, \citenamefont {Denisenko}, \citenamefont {Gelman},\ and\
  \citenamefont {Nori}}]{Satanin2014}%
  \BibitemOpen
  \bibfield  {author} {\bibinfo {author} {\bibfnamefont {A.~M.}\ \bibnamefont
  {Satanin}}, \bibinfo {author} {\bibfnamefont {M.~V.}\ \bibnamefont
  {Denisenko}}, \bibinfo {author} {\bibfnamefont {A.~I.}\ \bibnamefont
  {Gelman}},\ and\ \bibinfo {author} {\bibfnamefont {F.}~\bibnamefont {Nori}},\
  }\href@noop {} {\bibfield  {journal} {\bibinfo  {journal} {Phys. Rev. B}\
  }\textbf {\bibinfo {volume} {90}},\ \bibinfo {pages} {104516} (\bibinfo
  {year} {2014})}\BibitemShut {NoStop}%
\end{thebibliography}%

\end{document}